\documentclass[graybox]{svmult}

\usepackage[utf8]{inputenc}
\usepackage[T1]{fontenc}
\usepackage{type1cm}        
\usepackage{makeidx}         
\usepackage{graphicx}        
\usepackage{multicol}        
\usepackage[bottom]{footmisc}

\usepackage{newtxtext}       %
\usepackage{newtxmath}       

\usepackage{amsmath}
\usepackage{amsfonts}
\usepackage{accents}
\usepackage[colorlinks=true]{hyperref}

\makeatletter
\providecommand*{\toclevel@title}{0}
\providecommand*{\toclevel@author}{0}
\providecommand*{\toclevel@titlech}{0}
\providecommand*{\toclevel@authorch}{0}
\makeatother
\allowdisplaybreaks

\makeindex             

\newcommand{\order}[2]{\accentset{#2}{#1}}
\newcommand{\lc}[1]{\accentset{\circ}{#1}}
\newcommand{\dd}{\mathrm{d}}

\begin{document}

\title*{Teleparallel gravity}
\titlerunning{Teleparallel gravity}
\author{Manuel Hohmann}
\institute{Manuel Hohmann \at Laboratory of Theoretical Physics, Institute of Physics, University of Tartu, W. Ostwaldi 1, 50411 Tartu, Estonia \email{manuel.hohmann@ut.ee}}

\maketitle

\abstract{In general relativity, the only dynamical field describing the gravitational interaction of matter, is the metric. It induces the causal structure of spacetime, governs the motion of physical bodies through its Levi-Civita connection, and mediates gravity via the curvature of this connection. While numerous modified theories of gravity retain these principles, it is also possible to introduce another affine connection as a fundamental field, and consider its properties - curvature, torsion, nonmetricity - as the mediators of gravity. In the most general case, this gives rise to the class of metric-affine gravity theories, while restricting to metric-compatible connections, for which nonmetricity vanishes, comprises the class of Poincaré gauge theories. Alternatively, one may also consider connections with vanishing curvature. This assumption yields the class of \emph{teleparallel} gravity theories. This chapter gives a simplified introduction to teleparallel gravity, with a focus on performing practical calculations, as well as an overview of the most commonly studied classes of teleparallel gravity theories.}

\section{Introduction}\label{sec:intro}
In his original work, Einstein formulated the general theory of relativity in terms of the metric tensor as the fundamental field variable of the gravitational field, which describes gravity by the curvature of its Levi-Civita connection. Numerous modified gravity theories depart from this formulation, either keeping the metric as the only fundamental field variable and modifying its dynamics through a modified action, or by adding further fundamental field which couple non-minimally to the curvature~\cite{CANTATA:2021ktz}. However, there exist also other classes of gravity theories, in which the curvature of the Levi-Civita connection plays a less prominent role, and another, independent connection is introduced as a fundamental field variable next to the metric. Unlike the Levi-Civita connection, this connection is assumed to have vanishing curvature, but instead one allows for non-vanishing torsion or nonmetricity, or both. Gravity theories of this type are known as \emph{teleparallel} gravity theories.

In fact, already Einstein studied the possibility to describe gravity in terms of the torsion of a flat, metric-compatible connection instead of curvature~\cite{Einstein:1928}, in an attempt to unify gravity and electromagnetism. While this attempt was not successful, it gave rise to a new class of gravity theories, now known as metric teleparallel gravity theories~\cite{Aldrovandi:2013wha,Bahamonde:2021gfp}, in which gravity is mediated by torsion instead of curvature. Only much later another class of gravity theories was introduced, which attributes gravity to the nonmetricity of a flat, torsion-free (i.e., symmetric) connection, and is hence known as symmetric teleparallel gravity~\cite{Nester:1998mp}. Finally, allowing for both torsion and nonmetricity leads to the realm of general teleparallel gravity~\cite{BeltranJimenez:2019odq}. It is worth mentioning that these theories are embedded in the much wider and well-studied framework of metric-affine gravity theories~\cite{Hehl:1994ue,Blagojevic:2002du}, for the metric-compatible case also in the framework of Poincaré gauge theories~\cite{Hehl:1976kj,Blagojevic:2013xpa}. However, a full account of this relationship and the historic development and studies of teleparallel gravity theories would by far exceed the scope of this chapter.

Despite the long-standing history of teleparallel gravity theories and the studies of their fundamental properties and underlying structure for several decades, a renewed and growing interest in teleparallel modifications and extension of general relativity and their phenomenology has arisen only recently with the growing number of unexplained observations and tensions in cosmology. Numerous theories have been constructed as possible candidates to explain the early and late accelerating phases of the universe, known as inflation and dark energy eras, to resolve the question of singularities and the information paradox of black holes, and to provide alternative pathways towards a quantization of gravity and a unification with other fundamental forces. The phenomenology of these theories greatly differs and depends on their choice of dynamical fields and action, so that a full account would, again, exceed the scope of this chapter, and we must limit ourselves to a more general discussion of the class of teleparallel gravity theories, and leave specific theories and their phenomenological properties for further reading.

The aim of this chapter is to provide a practical introduction to teleparallel gravity. In section~\ref{sec:teleact} we give a simplified summary of the general structure and underlying mathematical foundations of teleparallel gravity theories in their three flavors - general, symmetric and metric. In particular, we discuss the fundamental fields in these theories, the general form of the action and the field equations. This practical introduction continues in section~\ref{sec:telephys}, where we explain how to formulate physical principles and perform common calculations necessary to solve the gravitational field equations of teleparallel gravity theories. We discuss how the invariance of the action under diffeomorphisms leads to the conservation of the matter currents, and show how to construct teleparallel geometries with spacetime symmetries and their perturbations, which can be used to solve the field equations of a given theory of gravity and thus study its phenomenology. Finally, section~\ref{sec:theories} gives an overview of the most commonly studied classes of teleparallel gravity theories and their field equations, and briefly summarizes their common properties.

There are many interesting aspects of teleparallel gravity which cannot be covered in this chapter, as they would by far exceed its scope and its aim towards performing practical calculations. In particular, we do not discuss the role of gauge symmetries in teleparallel gravity, which allow its interpretation as a gauge theory of the translation group. In relation to this, we do not discuss its formulation in terms of a tetrad. Throughout the chapter, we use only the tensor notation, which is more widespread in relating gravity to observations, and avoid the use of differential form language, which is often more concise and thus preferred by theorists, but less common in practical calculations of phenomenology. Further, we cannot cover fundamental questions such as the number of degrees of freedom of these theories, which is studied in their Hamiltonian formulation, and hints towards theoretical issues known under the term strong coupling. The interested reader is encouraged to follow the references provided in this chapter for a more detailed account of these mathematical foundations, their applications and possible issues.

We use the convention that spacetime indices are labeled with lowercase Greek letters and take the values \((0,1,2,3)\), as well as the metric signature \((-1,+1,+1,+1)\).

\section{Dynamical fields, action and field equations}\label{sec:teleact}
In this introductory section we give an overview of the dynamical fields and their properties, the general structure of the action, and the variational methods used to obtain their field equations. Here we focus on three different flavors of teleparallel theories: general teleparallel theories, in which both torsion and nonmetricity are allowed to be non-vanishing, are discussed in section~\ref{ssec:genteleact}; we then restrict the theories to symmetric teleparallel gravity by imposing vanishing torsion in section~\ref{sec:symteleact}, and to metric teleparallel gravity by imposing vanishing nonmetricity in section~\ref{sec:metteleact}.

\subsection{General teleparallel gravity}\label{ssec:genteleact}
We start our discussion of teleparallel gravity theories from the viewpoint of metric-affine gravity, in which next to the metric \(g_{\mu\nu}\) a connection with coefficients \(\Gamma^{\mu}{}_{\nu\rho}\) is introduced as a fundamental field on the spacetime manifold \(M\), which is independent of the Levi-Civita connection. To distinguish these two connections, we write the latter, and all derived quantities such as the covariant derivative and the curvature tensor, with an empty circle on top, i.e.,
\begin{equation}\label{eq:levicivita}
\lc{\Gamma}^{\mu}{}_{\nu\rho} = \frac{1}{2}g^{\mu\sigma}\left(\partial_{\nu}g_{\sigma\rho} + \partial_{\rho}g_{\nu\sigma} - \partial_{\sigma}g_{\nu\rho}\right)\,.
\end{equation}
Given another, independent connection, their difference can always be written in the form
\begin{equation}\label{eq:conndec}
\Gamma^{\mu}{}_{\nu\rho} - \lc{\Gamma}^{\mu}{}_{\nu\rho} = M^{\mu}{}_{\nu\rho} = K^{\mu}{}_{\nu\rho} + L^{\mu}{}_{\nu\rho}\,.
\end{equation}
Here, \(M^{\mu}{}_{\nu\rho}\) is called the \emph{distortion}: it is the difference between two connection coefficients, and hence a tensor field. If one of these two connections is the Levi-Civita connection of a metric, the distortion decomposes further into the \emph{contortion} \(K^{\mu}{}_{\nu\rho}\) and the \emph{disformation} \(L^{\mu}{}_{\nu\rho}\), which can be obtained as follows. First, define the \emph{torsion}
\begin{equation}\label{eq:torsion}
T^{\mu}{}_{\nu\rho} = \Gamma^{\mu}{}_{\rho\nu} - \Gamma^{\mu}{}_{\nu\rho}\,,
\end{equation}
as well as the \emph{nonmetricity}
\begin{equation}\label{eq:nonmetricity}
Q_{\mu\nu\rho} = \nabla_{\mu}g_{\nu\rho} = \partial_{\mu}g_{\nu\rho} - \Gamma^{\sigma}{}_{\nu\mu}g_{\sigma\rho} - \Gamma^{\sigma}{}_{\rho\mu}g_{\nu\sigma}\,.
\end{equation}
These are, again, tensor fields. Using the metric to raise and lower indices, one then obtains the contortion
\begin{equation}\label{eq:contortion}
K^{\mu}{}_{\nu\rho} = \frac{1}{2}\left(T_{\nu}{}^{\mu}{}_{\rho} + T_{\rho}{}^{\mu}{}_{\nu} - T^{\mu}{}_{\nu\rho}\right)\,,
\end{equation}
as well as the disformation
\begin{equation}\label{eq:disformation}
L^{\mu}{}_{\nu\rho} = \frac{1}{2}\left(Q^{\mu}{}_{\nu\rho} - Q_{\nu}{}^{\mu}{}_{\rho} - Q_{\rho}{}^{\mu}{}_{\nu}\right)\,.
\end{equation}
Hence, in the presence of a metric, an independent connection can always uniquely be specified in terms of its torsion and nonmetricity, which determine its deviation from the Levi-Civita connection.

The dynamical fields then enter the action of the theory, which is of the general form
\begin{equation}
S[g, \Gamma, \psi] = S_{\text{g}}[g, \Gamma] + S_{\text{m}}[g, \Gamma, \psi]\,,
\end{equation}
where the gravitational part \(S_{\text{g}}\) of the action depends only on the metric and the connection, while the matter part \(S_{\text{m}}\) also depends on some set of matter fields \(\psi^I\), whose components we do not specify further and simply label them with an index \(I\). By variation with respect to these matter fields, the matter action determines the matter field equations, which govern the dynamics of the matter fields in a given gravitational field background. In general, this background depends both on the metric \(g_{\mu\nu}\) and the connection \(\Gamma^{\mu}{}_{\nu\rho}\). Further, varying the matter action with respect to the metric and the connection gives rise to the \emph{energy-momentum} \(\Theta^{\mu\nu}\) and \emph{hypermomentum} \(H_{\mu}{}^{\nu\rho}\)\ defined by the variation~\cite{Hehl:1994ue}
\begin{equation}\label{eq:metricmatactvar}
\delta S_{\text{m}} = \int_M\left(\frac{1}{2}\Theta^{\mu\nu}\delta g_{\mu\nu} + H_{\mu}{}^{\nu\rho}\delta\Gamma^{\mu}{}_{\nu\rho} + \Psi_I\delta\psi^I\right)\sqrt{-g}\dd^4x\,,
\end{equation}
where \(\Psi_I = 0\) are the matter field equations. The specific form of \(\Theta^{\mu\nu}\) and \(H_{\mu}{}^{\nu\rho}\) depends on the type of matter under consideration and its coupling to the background geometry. These terms will act as the source of the gravitational field equations. To obtain the latter, one writes the variation of the gravitational part of the action in the similar form
\begin{equation}\label{eq:metricgravactvar}
\delta S_{\text{g}} = -\int_M\left(\frac{1}{2}W^{\mu\nu}\delta g_{\mu\nu} + Y_{\mu}{}^{\nu\rho}\delta\Gamma^{\mu}{}_{\nu\rho}\right)\sqrt{-g}\dd^4x\,,
\end{equation}
where any necessary integration by parts has been carried out in order to eliminate derivatives acting on the variations. This variation defines two further tensor fields, which we denote \(W^{\mu\nu}\) and \(Y_{\mu}{}^{\nu\rho}\), and which will enter as the dynamical part of the gravitational field equations.

The action and variation given above constitute the general form for a metric-affine theory of gravity. In teleparallel gravity, however, the connection is further restricted to have vanishing curvature,
\begin{equation}\label{eq:curvature}
R^{\mu}{}_{\nu\rho\sigma} = \partial_{\rho}\Gamma^{\mu}{}_{\nu\sigma} - \partial_{\sigma}\Gamma^{\mu}{}_{\nu\rho} + \Gamma^{\mu}{}_{\tau\rho}\Gamma^{\tau}{}_{\nu\sigma} - \Gamma^{\mu}{}_{\tau\sigma}\Gamma^{\tau}{}_{\nu\rho} \equiv 0\,.
\end{equation}
Note that this condition involves both the connection coefficients and their derivatives. In the context of Lagrange theory, such type of condition constitutes a nonholonomic constraint. Different possibilities exist to implement this constraint~\cite{Hohmann:2021fpr}. One possibility is to add another term of the form
\begin{equation}
S_{\text{r}} = \int_M\tilde{r}_{\mu}{}^{\nu\rho\sigma}R^{\mu}{}_{\nu\rho\sigma}\dd^4x\,,
\end{equation}
where the tensor density \(\tilde{r}_{\mu}{}^{\nu\rho\sigma}\) acts as a Lagrange multiplier, and can be taken to be antisymmetric in its last two indices, \(\tilde{r}_{\mu}{}^{\nu\rho\sigma} = \tilde{r}_{\mu}{}^{\nu[\rho\sigma]}\), since the contraction of its symmetric part with the antisymmetric indices of the curvature tensor vanishes and thus does not contribute to the action. Variation with respect to \(\tilde{r}_{\mu}{}^{\nu\rho\sigma}\) then yields the constraint equation \(R^{\mu}{}_{\nu\rho\sigma} = 0\). In order to derive the variation with respect to the connection coefficients, note that the variation of the curvature can be expressed as
\begin{equation}\label{eq:metaffvarflat}
\delta R^{\mu}{}_{\nu\rho\sigma} = \nabla_{\rho}\delta\Gamma^{\mu}{}_{\nu\sigma} - \nabla_{\sigma}\delta\Gamma^{\mu}{}_{\nu\rho} + T^{\tau}{}_{\rho\sigma}\delta\Gamma^{\mu}{}_{\nu\tau}\,.
\end{equation}
With the help of this expression, as well as performing integration by parts, one obtains the variation of the Lagrange multiplier term \(S_{\text{r}}\) in the action with respect to the connection as
\begin{equation}
\begin{split}
\delta_{\Gamma}S_{\text{r}} &= \int_M\tilde{r}_{\mu}{}^{\nu\rho\sigma}\left(\nabla_{\rho}\delta\Gamma^{\mu}{}_{\nu\sigma} - \nabla_{\sigma}\delta\Gamma^{\mu}{}_{\nu\rho} + T^{\tau}{}_{\rho\sigma}\delta\Gamma^{\mu}{}_{\nu\tau}\right)\dd^4x\\
&= \int_M\left(T^{\sigma}{}_{\sigma\rho}\tilde{r}_{\mu}{}^{\nu\rho\tau} - T^{\rho}{}_{\rho\sigma}\tilde{r}_{\mu}{}^{\nu\tau\sigma} + T^{\tau}{}_{\rho\sigma}\tilde{r}_{\mu}{}^{\nu\rho\sigma} - \nabla_{\rho}\tilde{r}_{\mu}{}^{\nu\rho\tau} + \nabla_{\sigma}\tilde{r}_{\mu}{}^{\nu\tau\sigma}\right)\delta\Gamma^{\mu}{}_{\nu\tau}\dd^4x\,.
\end{split}
\end{equation}
Combining all terms, one finds that the gravitational field equations are given by the metric field equation
\begin{equation}\label{eq:genmetfield}
W_{\mu\nu} = \Theta_{\mu\nu}\,,
\end{equation}
as well as the connection field equation
\begin{equation}\label{eq:genconnfieldlag}
\tilde{Y}_{\mu}{}^{\nu\tau} = \tilde{H}_{\mu}{}^{\nu\tau} + T^{\sigma}{}_{\sigma\rho}\tilde{r}_{\mu}{}^{\nu\rho\tau} - T^{\rho}{}_{\rho\sigma}\tilde{r}_{\mu}{}^{\nu\tau\sigma} + T^{\tau}{}_{\rho\sigma}\tilde{r}_{\mu}{}^{\nu\rho\sigma} - \nabla_{\rho}\tilde{r}_{\mu}{}^{\nu\rho\tau} + \nabla_{\sigma}\tilde{r}_{\mu}{}^{\nu\tau\sigma}\,,
\end{equation}
where it is convenient to define the tensor densities
\begin{equation}\label{eq:connvardens}
\tilde{Y}_{\mu}{}^{\nu\tau} = Y_{\mu}{}^{\nu\tau}\sqrt{-g}\,, \quad
\tilde{H}_{\mu}{}^{\nu\tau} = H_{\mu}{}^{\nu\tau}\sqrt{-g}\,.
\end{equation}
Note that the connection still contains the undetermined Lagrange multiplier \(\tilde{r}_{\mu}{}^{\nu\rho\sigma}\). However, the latter can be eliminated using the following procedure. First, we calculate the divergence
\begin{multline}\label{eq:divconnfield}
\nabla_{\tau}\tilde{Y}_{\mu}{}^{\nu\tau} = \nabla_{\tau}\tilde{H}_{\mu}{}^{\nu\tau} + \nabla_{\tau}\left(T^{\sigma}{}_{\sigma\rho}\tilde{r}_{\mu}{}^{\nu\rho\tau} - T^{\rho}{}_{\rho\sigma}\tilde{r}_{\mu}{}^{\nu\tau\sigma} + T^{\tau}{}_{\rho\sigma}\tilde{r}_{\mu}{}^{\nu\rho\sigma}\right)\\
- \nabla_{\tau}\nabla_{\rho}\tilde{r}_{\mu}{}^{\nu\rho\tau} + \nabla_{\tau}\nabla_{\sigma}\tilde{r}_{\mu}{}^{\nu\tau\sigma}\,.
\end{multline}
The last two terms can be simplified by realizing that the Lagrange multiplier \(\tilde{r}_{\mu}{}^{\nu\rho\sigma}\) is antisymmetric in its last two indices, so that one can apply the commutator of covariant derivatives given by
\begin{multline}
2\nabla_{[\rho}\nabla_{\sigma]}\tilde{r}_{\mu}{}^{\nu\rho\sigma} = -T^{\tau}{}_{\rho\sigma}\nabla_{\tau}\tilde{r}_{\mu}{}^{\nu\rho\sigma}\\
- R^{\tau}{}_{\mu\rho\sigma}\tilde{r}_{\tau}{}^{\nu\rho\sigma} + R^{\nu}{}_{\tau\rho\sigma}\tilde{r}_{\mu}{}^{\tau\rho\sigma} + R^{\rho}{}_{\tau\rho\sigma}\tilde{r}_{\mu}{}^{\nu\tau\sigma} + R^{\sigma}{}_{\tau\rho\sigma}\tilde{r}_{\mu}{}^{\nu\rho\tau} - R^{\tau}{}_{\tau\rho\sigma}\tilde{r}_{\mu}{}^{\nu\rho\sigma}\,.
\end{multline}
Also using the vanishing curvature~\eqref{eq:curvature}, the only remaining term is given by
\begin{equation}
2\nabla_{[\rho}\nabla_{\sigma]}\tilde{r}_{\mu}{}^{\nu\rho\sigma} = -T^{\tau}{}_{\rho\sigma}\nabla_{\tau}\tilde{r}_{\mu}{}^{\nu\rho\sigma}\,.
\end{equation}
Further, one can use the antisymmetry of the Lagrange multiplier to write
\begin{multline}
\nabla_{\tau}\left(T^{\sigma}{}_{\sigma\rho}\tilde{r}_{\mu}{}^{\nu\rho\tau} - T^{\rho}{}_{\rho\sigma}\tilde{r}_{\mu}{}^{\nu\tau\sigma} + T^{\tau}{}_{\rho\sigma}\tilde{r}_{\mu}{}^{\nu\rho\sigma}\right) = 3\nabla_{[\tau}\left(T^{\tau}{}_{\rho\sigma]}\tilde{r}_{\mu}{}^{\nu\rho\sigma}\right).
\end{multline}
The derivative of the torsion tensor can be rewritten by making use of the curvature-free Bianchi identity
\begin{equation}
\nabla_{[\nu}T^{\mu}{}_{\rho\sigma]} + T^{\mu}{}_{\tau[\nu}T^{\tau}{}_{\rho\sigma]} = R^{\mu}{}_{[\nu\rho\sigma]} = 0\,,
\end{equation}
from which after contraction follows
\begin{equation}
3\nabla_{[\tau}T^{\tau}{}_{\rho\sigma]} = -3T^{\tau}{}_{\omega[\tau}T^{\omega}{}_{\rho\sigma]} = T^{\tau}{}_{\tau\omega}T^{\omega}{}_{\rho\sigma}\,.
\end{equation}
By combining all terms, one finds that the divergence~\eqref{eq:divconnfield} of the connection field equation~\eqref{eq:genconnfieldlag} reads
\begin{equation}
\begin{split}
\nabla_{\tau}\tilde{Y}_{\mu}{}^{\nu\tau} &= \nabla_{\tau}\tilde{H}_{\mu}{}^{\nu\tau} + T^{\tau}{}_{\tau\omega}T^{\omega}{}_{\rho\sigma}\tilde{r}_{\mu}{}^{\nu\rho\sigma} + 3T^{\tau}{}_{[\rho\sigma}\nabla_{\tau]}\tilde{r}_{\mu}{}^{\nu\rho\sigma} - T^{\tau}{}_{\rho\sigma}\nabla_{\tau}\tilde{r}_{\mu}{}^{\nu\rho\sigma}\\
&= \nabla_{\tau}\tilde{H}_{\mu}{}^{\nu\tau} + T^{\tau}{}_{\tau\omega}T^{\omega}{}_{\rho\sigma}\tilde{r}_{\mu}{}^{\nu\rho\sigma} + 2T^{\tau}{}_{\tau[\rho}\nabla_{\sigma]}\tilde{r}_{\mu}{}^{\nu\rho\sigma}\,.
\end{split}
\end{equation}
Similarly, contracting the field equation~\eqref{eq:genconnfieldlag} with the trace of the torsion tensor, one obtains
\begin{equation}
T^{\omega}{}_{\omega\tau}\tilde{Y}_{\mu}{}^{\nu\tau} = T^{\omega}{}_{\omega\tau}\tilde{H}_{\mu}{}^{\nu\tau} + T^{\omega}{}_{\omega\tau}T^{\tau}{}_{\rho\sigma}\tilde{r}_{\mu}{}^{\nu\rho\sigma} + 2T^{\omega}{}_{\omega[\rho}\nabla_{\sigma]}\tilde{r}_{\mu}{}^{\nu\rho\sigma}\,.
\end{equation}
Subtracting these two equations, the Lagrange multiplier terms cancel, and one obtains the connection field equations
\begin{equation}\label{eq:genconnfielddens}
\nabla_{\tau}\tilde{Y}_{\mu}{}^{\nu\tau} - T^{\omega}{}_{\omega\tau}\tilde{Y}_{\mu}{}^{\nu\tau} = \nabla_{\tau}\tilde{H}_{\mu}{}^{\nu\tau} - T^{\omega}{}_{\omega\tau}\tilde{H}_{\mu}{}^{\nu\tau}\,.
\end{equation}
This equation can also be rewritten by eliminating the density factors using
\begin{equation}\label{eq:covderdens}
\nabla_{\mu}\sqrt{-g} = \frac{1}{2}g^{\nu\rho}\nabla_{\mu}g_{\nu\rho}\sqrt{-g} = \frac{1}{2}Q_{\mu\nu}{}^{\nu}\sqrt{-g} = M^{\nu}{}_{\nu\mu}\sqrt{-g}\,,
\end{equation}
where the last expression follows from rewriting the covariant derivative of the metric in terms of its (vanishing) covariant derivative with respect to the Levi-Civita connection using the decomposition~\eqref{eq:conndec}. The latter can also be used to write the torsion as
\begin{equation}
T^{\mu}{}_{\nu\rho} = M^{\mu}{}_{\rho\nu} - M^{\mu}{}_{\nu\rho}\,.
\end{equation}
Using these relations and the definition~\eqref{eq:connvardens} of the densities \(\tilde{Y}_{\mu}{}^{\nu\rho}\) and \(\tilde{H}_{\mu}{}^{\nu\rho}\), the field equations become
\begin{equation}\label{eq:genconnfield}
\nabla_{\tau}Y_{\mu}{}^{\nu\tau} - M^{\omega}{}_{\tau\omega}Y_{\mu}{}^{\nu\tau} = \nabla_{\tau}H_{\mu}{}^{\nu\tau} - M^{\omega}{}_{\tau\omega}H_{\mu}{}^{\nu\tau}\,.
\end{equation}
The equations~\eqref{eq:genmetfield} and~\eqref{eq:genconnfield} constitute the field equations for the dynamical fields in teleparallel gravity.

Besides the method of Lagrange multipliers, the teleparallel field equations can also be obtained by using the method of restricted variation. Using this method, no Lagrange multiplier is introduced, the constraint equation~\eqref{eq:curvature} of vanishing curvature is imposed to restrict the connection \(\Gamma^{\mu}{}_{\nu\rho}\), and the variation \(\delta\Gamma^{\mu}{}_{\nu\rho}\) is restricted in order to preserve this constraint. Using the expression~\eqref{eq:metaffvarflat}, one finds that the variation of the connection must be of the form
\begin{equation}\label{eq:flatvar}
\delta\Gamma^{\mu}{}_{\nu\rho} = \nabla_{\rho}\xi^{\mu}{}_{\nu}
\end{equation}
for a tensor field \(\xi^{\mu}{}_{\nu}\). Indeed, for the curvature perturbation one then finds
\begin{equation}
\delta R^{\mu}{}_{\nu\rho\sigma} = \nabla_{\rho}\nabla_{\sigma}\xi^{\mu}{}_{\nu} - \nabla_{\sigma}\nabla_{\rho}\xi^{\mu}{}_{\nu} + T^{\tau}{}_{\rho\sigma}\nabla_{\tau}\xi^{\mu}{}_{\nu} = 0\,,
\end{equation}
using the formula for the commutator of covariant derivatives in the absence of curvature. It follows that the variation of the action takes the form
\begin{equation}
\begin{split}
\delta_{\Gamma}S &= \int_M\left(\tilde{H}_{\mu}{}^{\nu\rho} - \tilde{Y}_{\mu}{}^{\nu\rho}\right)\nabla_{\rho}\xi^{\mu}{}_{\nu}\dd^4x\\
&= \int_M\left(T^{\sigma}{}_{\sigma\rho}\tilde{H}_{\mu}{}^{\nu\rho} - \nabla_{\rho}\tilde{H}_{\mu}{}^{\nu\rho} - T^{\sigma}{}_{\sigma\rho}\tilde{Y}_{\mu}{}^{\nu\rho} + \nabla_{\rho}\tilde{Y}_{\mu}{}^{\nu\rho}\right)\xi^{\mu}{}_{\nu}\dd^4x\,,
\end{split}
\end{equation}
where the second line follows from integration by parts. Hence, one finds the same connection field equation~\eqref{eq:genconnfielddens}.

\subsection{Symmetric teleparallel gravity}\label{sec:symteleact}
The class of teleparallel gravity theories discussed in the previous section, in which the affine connection \(\Gamma^{\mu}{}_{\nu\rho}\) is restricted only by the flatness condition~\eqref{eq:curvature}, is also known as \emph{general} teleparallel gravity, and is the youngest among the different classes of teleparallel gravity theories. Two other classes of teleparallel gravity theories can be obtained by demanding that either the torsion~\eqref{eq:torsion} or the nonmetricity~\eqref{eq:nonmetricity} vanishes. We will start with the former condition, which yields the class of \emph{symmetric} teleparallel gravity theories, which refers to the fact that the coefficients of a torsion-free connection are symmetric in their lower two indices. In order to implement the condition of vanishing torsion, one may proceed in full analogy to the flatness condition in the previous section, by adding another Lagrange multiplier term
\begin{equation}\label{eq:lagmultors}
S_{\text{t}} = \int_M\tilde{t}_{\mu}{}^{\nu\rho}T^{\mu}{}_{\nu\rho}\dd^4x\,,
\end{equation}
where variation with respect to the tensor density \(\tilde{t}_{\mu}{}^{\nu\rho}\) leads to the constraint equation \(T^{\mu}{}_{\nu\rho} = 0\). In order to derive the field equations, one then proceeds as in the previous section, by varying the full action and eliminating the Lagrange multipliers from the resulting field equations. This calculation is rather lengthy, but straightforward, and so we will not show it here. Instead, we will follow the alternative procedure of restricted variation of the action, by considering only variations \(\delta\Gamma^{\mu}{}_{\nu\rho}\) which maintain the vanishing curvature and torsion of the connection. We can use the fact that the flatness is maintained by the variation~\eqref{eq:flatvar}, and further restrict the form of \(\xi^{\mu}{}_{\nu}\). It turns out that this is achieved by setting \(\xi^{\mu}{}_{\nu} = \nabla_{\nu}\zeta^{\mu}\) for some vector field \(\zeta^{\mu}\), and thus
\begin{equation}\label{eq:symtelevar}
\delta\Gamma^{\mu}{}_{\nu\rho} = \nabla_{\rho}\nabla_{\nu}\zeta^{\mu}\,.
\end{equation}
Using the fact that covariant derivatives commute in the absence of curvature and torsion, one now immediately sees
\begin{equation}
\delta T^{\mu}{}_{\nu\rho} = \delta\Gamma^{\mu}{}_{\rho\nu} - \delta\Gamma^{\mu}{}_{\nu\rho} = \nabla_{\nu}\nabla_{\rho}\zeta^{\mu} - \nabla_{\rho}\nabla_{\nu}\zeta^{\mu} = 0\,.
\end{equation}
The variation of the action with respect to the connection is then simply given by
\begin{equation}
\begin{split}
\delta_{\Gamma}S &= \int_M\left(\tilde{H}_{\mu}{}^{\nu\rho} - \tilde{Y}_{\mu}{}^{\nu\rho}\right)\nabla_{\rho}\nabla_{\nu}\zeta^{\mu}\dd^4x\\
&= -\int_M\nabla_{\rho}\left(\tilde{H}_{\mu}{}^{\nu\rho} - \tilde{Y}_{\mu}{}^{\nu\rho}\right)\nabla_{\nu}\zeta^{\mu}\dd^4x\\
&= \int_M\nabla_{\nu}\nabla_{\rho}\left(\tilde{H}_{\mu}{}^{\nu\rho} - \tilde{Y}_{\mu}{}^{\nu\rho}\right)\zeta^{\mu}\dd^4x\,,
\end{split}
\end{equation}
where integration by parts simplifies due to the vanishing torsion. The connection field equation thus becomes
\begin{equation}\label{eq:symconnfield}
\nabla_{\nu}\nabla_{\rho}\tilde{Y}_{\mu}{}^{\nu\rho} = \nabla_{\nu}\nabla_{\rho}\tilde{H}_{\mu}{}^{\nu\rho}\,.
\end{equation}
Together with the metric field equation~\eqref{eq:genmetfield}, it constitutes the field equations of symmetric teleparallel gravity.

\subsection{Metric teleparallel gravity}\label{sec:metteleact}
We finally come to the remaining class of theories, which are defined by imposing the condition of vanishing nonmetricity, so that the connection becomes metric-compatible. This class of theories is therefore known as \emph{metric} teleparallel gravity, or simply as teleparallel gravity, since it was conceived first among the three different classes we discuss here. To derive its field equations, one can also in this case either introduce a Lagrange multiplier
\begin{equation}\label{eq:lagmulnonmet}
S_{\text{q}} = \int_M\tilde{q}^{\mu\nu\rho}Q_{\mu\nu\rho}\dd^4x\,,
\end{equation}
and vary with respect to the tensor density \(\tilde{q}^{\mu\nu\rho}\) to obtain \(Q_{\mu\nu\rho} = 0\), or find a suitable restriction on the connection variation. Here we will follow once again the latter approach. From the definition~\eqref{eq:nonmetricity} of the nonmetricity, one obtains its variation
\begin{equation}
\delta Q_{\mu\nu\rho} = \nabla_{\mu}\delta g_{\nu\rho} - g_{\sigma\rho}\delta\Gamma^{\sigma}{}_{\nu\mu} - g_{\nu\sigma}\delta\Gamma^{\sigma}{}_{\rho\mu} = \nabla_{\mu}(\delta g_{\nu\rho} - 2\xi_{(\nu\rho)})\,,
\end{equation}
provided that the variation of the connection is chosen to implement the flatness condition~\eqref{eq:flatvar}. Here we also used the metric compatibility of the connection to commute lowering an index with the covariant derivative. It turns out that the condition of vanishing nonmetricity imposes a relation
\begin{equation}
\delta g_{\mu\nu} = 2\xi_{(\mu\nu)}
\end{equation}
between the variations of the metric and the connection. Since both are now expressed in terms of the tensor field \(\xi_{\mu\nu}\), the field equations follow from the total variation
\begin{equation}
\begin{split}
\delta S &= \int_M\left(\Theta^{\mu\nu}\xi_{(\mu\nu)} + H^{\mu\nu\rho}\nabla_{\rho}\xi_{\mu\nu} - W^{\mu\nu}\xi_{(\mu\nu)} - Y^{\mu\nu\rho}\nabla_{\rho}\xi_{\mu\nu}\right)\sqrt{-g}\dd^4x\\
&= \int_M\left(\Theta^{(\mu\nu)} - \nabla_{\rho}H^{\mu\nu\rho} + H^{\mu\nu\rho}T^{\tau}{}_{\tau\rho} - W^{(\mu\nu)} + \nabla_{\rho}Y^{\mu\nu\rho} - Y^{\mu\nu\rho}T^{\tau}{}_{\tau\rho}\right)\xi_{\mu\nu}\sqrt{-g}\dd^4x\,,
\end{split}
\end{equation}
after performing integration by parts, and using the metric compatibility of the connection to obtain \(\nabla_{\mu}\sqrt{-g} = 0\). Keeping in mind that \(W^{\mu\nu}\) and \(\Theta^{\mu\nu}\) are defined by the variation of the action with respect to the metric, and thus symmetric by definition, one obtains the field equation
\begin{equation}\label{eq:metallfield}
W^{\mu\nu} - \nabla_{\rho}Y^{\mu\nu\rho} + Y^{\mu\nu\rho}T^{\tau}{}_{\tau\rho} = \Theta^{\mu\nu} - \nabla_{\rho}H^{\mu\nu\rho} + H^{\mu\nu\rho}T^{\tau}{}_{\tau\rho}\,.
\end{equation}
This single field equation therefore conveys the dynamics in metric teleparallel gravity.

\section{Physical aspects and formalisms in teleparallel geometry}\label{sec:telephys}
To be able to make contact with phenomenology and observations, it is necessary to discuss a few general physical principles in the framework of teleparallel gravity. The first principle, which we discuss in section~\ref{ssec:enmomhypcons}, is the conservation of the matter currents, which are energy-momentum and hypermomentum, which follows from the invariance of the action under diffeomorphisms. We then continue with spacetime symmetries in section~\ref{ssec:symmetry}, which can be used to obtain solutions of teleparallel gravity theories, such as black holes, whose phenomenology can subsequently be studied. In particular, we focus on the case of homogeneous and isotropic teleparallel spacetimes, and derive the dynamical variables which appear in teleparallel cosmology. Finally, we discuss the theory of perturbations of teleparallel geometries in section~\ref{ssec:pert}. These form the basis of testing teleparallel gravity theories using gravitational waves and high-precision post-Newtonian observations.

\subsection{Energy-momentum-hypermomentum conservation}\label{ssec:enmomhypcons}
In order to be independent of the choice of coordinates, the different components \(S_{\text{g}}\) and \(S_{\text{m}}\) of the action discussed in the previous sections are demanded to be independently invariant under diffeomorphisms. Note that an infinitesimal diffeomorphism generated by a vector field \(X^{\mu}\) changes the metric by
\begin{equation}\label{eq:mettrans}
\delta_Xg_{\mu\nu} = (\mathcal{L}_{X}g)_{\mu\nu} = X^{\rho}\partial_{\rho}g_{\mu\nu} + \partial_{\mu}X^{\rho}g_{\rho\nu} + \partial_{\nu}X^{\rho}g_{\mu\rho} = 2\lc{\nabla}_{(\mu}X_{\nu)}\,,
\end{equation}
while the connection is changed by
\begin{equation}\label{eq:conntrans}
\begin{split}
\delta_{X}\Gamma^{\mu}{}_{\nu\rho} &= (\mathcal{L}_{X}\Gamma)^{\mu}{}_{\nu\rho}\\
&= X^{\sigma}\partial_{\sigma}\Gamma^{\mu}{}_{\nu\rho} - \partial_{\sigma}X^{\mu}\Gamma^{\sigma}{}_{\nu\rho} + \partial_{\nu}X^{\sigma}\Gamma^{\mu}{}_{\sigma\rho} + \partial_{\rho}X^{\sigma}\Gamma^{\mu}{}_{\nu\sigma} + \partial_{\nu}\partial_{\rho}X^{\mu}\\
&= \nabla_{\rho}\nabla_{\nu}X^{\mu} - X^{\sigma}R^{\mu}{}_{\nu\rho\sigma} - \nabla_{\rho}(X^{\sigma}T^{\mu}{}_{\nu\sigma})\,.
\end{split}
\end{equation}
Note that both expressions are tensor fields, despite the fact that the connection coefficients are not tensor fields. Their variation, however, being an infinitesimal difference between connection coefficients, is a tensor field. In the teleparallel case, the curvature tensor vanishes. Using these formulas, it is now easy to calculate the change of the gravitational part \(S_{\text{g}}\) of the action, which reads
\begin{equation}
\begin{split}
\delta_XS_{\text{g}} &= -\int_M\left(\frac{1}{2}\sqrt{-g}W^{\mu\nu}\delta_Xg_{\mu\nu} + \tilde{Y}_{\mu}{}^{\nu\rho}\delta_X\Gamma^{\mu}{}_{\nu\rho}\right)\dd^4x\\
&= -\int_M\left\{\sqrt{-g}W^{\mu\nu}\lc{\nabla}_{\mu}X_{\nu} + \tilde{Y}_{\mu}{}^{\nu\rho}\left[\nabla_{\rho}\nabla_{\nu}X^{\mu} - \nabla_{\rho}(X^{\sigma}T^{\mu}{}_{\nu\sigma})\right]\right\}\dd^4x\\
&= \int_M\Big[\sqrt{-g}\lc{\nabla}_{\nu}W_{\mu}{}^{\nu} + T^{\sigma}{}_{\mu\nu}(\nabla_{\rho}\tilde{Y}_{\sigma}{}^{\nu\rho} - T^{\tau}{}_{\tau\rho}\tilde{Y}_{\sigma}{}^{\nu\rho})\\
&\phantom{=}- \nabla_{\nu}(\nabla_{\rho}\tilde{Y}_{\mu}{}^{\nu\rho} - T^{\tau}{}_{\tau\rho}\tilde{Y}_{\mu}{}^{\nu\rho}) + T^{\omega}{}_{\omega\nu}(\nabla_{\rho}\tilde{Y}_{\mu}{}^{\nu\rho} - T^{\tau}{}_{\tau\rho}\tilde{Y}_{\mu}{}^{\nu\rho})\Big]X^{\mu}\dd^4x\,.
\end{split}
\end{equation}
Assuming that the gravitational part \(S_{\text{g}}\) of the action is invariant under diffeomorphisms, this variation must vanish identically for arbitrary vector fields \(X^{\mu}\). Hence, it follows that the terms \(W^{\mu\nu}\) and \(\tilde{Y}_{\mu}{}^{\nu\rho}\) obtained from the variation of the action satisfy
\begin{multline}\label{eq:gravconsdens}
\sqrt{-g}\lc{\nabla}_{\nu}W_{\mu}{}^{\nu} + T^{\sigma}{}_{\mu\nu}(\nabla_{\rho}\tilde{Y}_{\sigma}{}^{\nu\rho} - T^{\tau}{}_{\tau\rho}\tilde{Y}_{\sigma}{}^{\nu\rho})\\
- \nabla_{\nu}(\nabla_{\rho}\tilde{Y}_{\mu}{}^{\nu\rho} - T^{\tau}{}_{\tau\rho}\tilde{Y}_{\mu}{}^{\nu\rho}) + T^{\sigma}{}_{\sigma\nu}(\nabla_{\rho}\tilde{Y}_{\mu}{}^{\nu\rho} - T^{\tau}{}_{\tau\rho}\tilde{Y}_{\mu}{}^{\nu\rho}) = 0\,.
\end{multline}
Alternatively, one can also write this relation without density factors, and finds
\begin{multline}\label{eq:gravcons}
\lc{\nabla}_{\nu}W_{\mu}{}^{\nu} + T^{\sigma}{}_{\mu\nu}(\nabla_{\rho}Y_{\sigma}{}^{\nu\rho} - M^{\tau}{}_{\rho\tau}Y_{\sigma}{}^{\nu\rho})\\
- \nabla_{\nu}(\nabla_{\rho}Y_{\mu}{}^{\nu\rho} - M^{\tau}{}_{\rho\tau}Y_{\mu}{}^{\nu\rho}) + M^{\sigma}{}_{\nu\sigma}(\nabla_{\rho}Y_{\mu}{}^{\nu\rho} - M^{\tau}{}_{\rho\tau}Y_{\mu}{}^{\nu\rho}) = 0\,.
\end{multline}
This equation is derived from a purely geometric property of the gravitational part of the action, and so it is a geometric identity, i.e., it holds for any field configuration of the metric \(g_{\mu\nu}\) and the connection \(\Gamma^{\mu}{}_{\nu\rho}\), independently of whether these satisfy the gravitational field equations or not. Such a relation is therefore also said to hold \emph{off-shell}. This is to be contrasted with the variation of the matter action \(S_{\text{m}}\), which reads
\begin{equation}
\begin{split}
\delta_XS_{\text{g}} &= \int_M\left(\frac{1}{2}\sqrt{-g}\Theta^{\mu\nu}\delta_Xg_{\mu\nu} + \tilde{H}_{\mu}{}^{\nu\rho}\delta_X\Gamma^{\mu}{}_{\nu\rho} + \tilde{\Psi}_I\delta_X\psi^I\right)\dd^4x\\
&= \int_M\left\{\sqrt{-g}\Theta^{\mu\nu}\lc{\nabla}_{\mu}X_{\nu} + \tilde{H}_{\mu}{}^{\nu\rho}\left[\nabla_{\rho}\nabla_{\nu}X^{\mu} - \nabla_{\rho}(X^{\sigma}T^{\mu}{}_{\nu\sigma})\right] + \tilde{\Psi}_I\mathcal{L}_X\psi^I\right\}\dd^4x\,.
\end{split}
\end{equation}
Here, \(\tilde{\Psi}_I = 0\) (or equivalently \(\Psi_I = 0\), without using densities) are the matter field equations. If these are satisfied, and only then, demanding that the matter action is invariant under diffeomorphisms generated by an arbitrary vector field \(X^{\mu}\) leads to the energy-momentum-hypermomentum conservation law
\begin{multline}\label{eq:enmomhypconsdens}
\sqrt{-g}\lc{\nabla}_{\nu}\Theta_{\mu}{}^{\nu} + T^{\sigma}{}_{\mu\nu}(\nabla_{\rho}\tilde{H}_{\sigma}{}^{\nu\rho} - T^{\tau}{}_{\tau\rho}\tilde{H}_{\sigma}{}^{\nu\rho})\\
- \nabla_{\nu}(\nabla_{\rho}\tilde{H}_{\mu}{}^{\nu\rho} - T^{\tau}{}_{\tau\rho}\tilde{H}_{\mu}{}^{\nu\rho}) + T^{\sigma}{}_{\sigma\nu}(\nabla_{\rho}\tilde{H}_{\mu}{}^{\nu\rho} - T^{\tau}{}_{\tau\rho}\tilde{H}_{\mu}{}^{\nu\rho}) = 0\,,
\end{multline}
or, again in the version without densities,
\begin{multline}\label{eq:enmomhypcons}
\lc{\nabla}_{\nu}\Theta_{\mu}{}^{\nu} + T^{\sigma}{}_{\mu\nu}(\nabla_{\rho}H_{\sigma}{}^{\nu\rho} - M^{\tau}{}_{\rho\tau}H_{\sigma}{}^{\nu\rho})\\
- \nabla_{\nu}(\nabla_{\rho}H_{\mu}{}^{\nu\rho} - M^{\tau}{}_{\rho\tau}H_{\mu}{}^{\nu\rho}) + M^{\sigma}{}_{\nu\sigma}(\nabla_{\rho}H_{\mu}{}^{\nu\rho} - M^{\tau}{}_{\rho\tau}H_{\mu}{}^{\nu\rho}) = 0\,.
\end{multline}
Since this relation does not hold for arbitrary field configurations of the gravitational and matter field, but only for those which satisfy the matter field equations \(\Psi_I = 0\), it is said to hold \emph{on-shell}. Note that we have not made any assumptions on the properties of the connection except for vanishing curvature. In particular, we have not imposed vanishing torsion or nonmetricity. It follows that the geometric identity and energy-momentum-hypermomentum law given above hold for all three classes of teleparallel gravity theories (but their expressions will simplify in the symmetric and metric cases, as we will see below). Finally, we remark that in the case of vanishing hypermomentum, i.e., for matter which couples only to the metric and not to the connection, which is most commonly considered in the context of teleparallel gravity, the conservation law reduces to
\begin{equation}
\lc{\nabla}_{\nu}\Theta_{\mu}{}^{\nu} = 0\,,
\end{equation}
which is the well-known energy-momentum conservation.

As an alternative to imposing the matter field equations, the conservation law~\eqref{eq:enmomhypcons} can also be derived from the geometric identity~\eqref{eq:gravcons}, by imposing the gravitational field equations. This is most straightforward for the general teleparallel gravity class, whose gravitational field equations are~\eqref{eq:genmetfield} and~\eqref{eq:genconnfield}. One easily sees that the terms appearing in the identity~\eqref{eq:gravcons} are exactly the left-hand sides of the gravitational field equations. Replacing them with the respective right-hand sides, one obtains the energy-momentum-hypermomentum conservation law~\eqref{eq:enmomhypcons}. Of course, the same holds true also if one uses the tensor density version of these equations.

A similar derivation can also be used in the case of symmetric teleparallel gravity, where one assumes vanishing torsion, \(T^{\mu}{}_{\nu\rho} = 0\). In this case, it is most convenient to start from the density version~\eqref{eq:gravconsdens}, which simplifies to become
\begin{equation}\label{eq:symgravcons}
\sqrt{-g}\lc{\nabla}_{\nu}W_{\mu}{}^{\nu} - \nabla_{\nu}\nabla_{\rho}\tilde{Y}_{\mu}{}^{\nu\rho} = 0\,.
\end{equation}
Using the metric field equation~\eqref{eq:genmetfield} and the connection field equation~\eqref{eq:symconnfield}, one thus immediately obtains the conservation law
\begin{equation}\label{eq:symenmomhypcons}
\sqrt{-g}\lc{\nabla}_{\nu}\Theta_{\mu}{}^{\nu} - \nabla_{\nu}\nabla_{\rho}\tilde{H}_{\mu}{}^{\nu\rho} = 0\,,
\end{equation}
which agrees with the general form~\eqref{eq:enmomhypconsdens} in the absence of torsion. What is most remarkable in the case of symmetric teleparallel gravity is the fact that one can also proceed in a different order: by imposing the matter field equations \(\Psi_I = 0\), from which follows the conservation law~\eqref{eq:symenmomhypcons}, further imposing the metric field equation~\eqref{eq:genmetfield}, and using the identity~\eqref{eq:symgravcons}, one obtains the connection field equation~\eqref{eq:symconnfield}. In other words, any field configuration of the matter and gravitational fields, which satisfies the matter and metric field equations, automatically satisfies also the connection field equation. For this reason, one often omits the latter when it comes to solving the field equations.

Finally, we study the energy-momentum-hypermomentum conservation also in the metric teleparallel setting. In this case, one can omit the density factors in the geometric identity~\eqref{eq:gravconsdens}, since the connection is metric-compatible, so that it becomes
\begin{multline}\label{eq:metgravcons}
\lc{\nabla}_{\nu}W_{\mu}{}^{\nu} + T^{\sigma}{}_{\mu\nu}(\nabla_{\rho}Y_{\sigma}{}^{\nu\rho} - T^{\tau}{}_{\tau\rho}Y_{\sigma}{}^{\nu\rho})\\
- \nabla_{\nu}(\nabla_{\rho}Y_{\mu}{}^{\nu\rho} - T^{\tau}{}_{\tau\rho}Y_{\mu}{}^{\nu\rho}) + T^{\sigma}{}_{\sigma\nu}(\nabla_{\rho}Y_{\mu}{}^{\nu\rho} - T^{\tau}{}_{\tau\rho}Y_{\mu}{}^{\nu\rho}) = 0\,.
\end{multline}
Further, we impose the metric teleparallel gravity field equation~\eqref{eq:metallfield}, which we will write in the form
\begin{equation}\label{eq:transmetfield}
W^{\mu\nu} - \Theta^{\mu\nu} = A^{\mu\nu}\,,
\end{equation}
where we have defined the abbreviation
\begin{equation}\label{eq:metfieldabbr}
A^{\mu\nu} = \nabla_{\rho}Y^{\mu\nu\rho} - Y^{\mu\nu\rho}T^{\tau}{}_{\tau\rho} - \nabla_{\rho}H^{\mu\nu\rho} + H^{\mu\nu\rho}T^{\tau}{}_{\tau\rho}\,.
\end{equation}
Note that the left hand side of the field equation~\eqref{eq:transmetfield} is symmetric by definition. Hence, when the equation holds, also the right hand side must be symmetric, and thus \(A^{[\mu\nu]} = 0\). We then take the Levi-Civita covariant derivative of this equation, which reads
\begin{equation}\label{eq:divmetfield}
\lc{\nabla}_{\nu}W^{\mu\nu} - \lc{\nabla}_{\nu}\Theta^{\mu\nu} = \lc{\nabla}_{\nu}A^{\mu\nu}\,.
\end{equation}
On the right-hand side, we can use the relation
\begin{equation}\label{eq:covdevmetfield}
\begin{split}
\lc{\nabla}_{\nu}A^{\mu\nu} &= \nabla_{\nu}A^{\mu\nu} - K^{\mu}{}_{\rho\nu}A^{\rho\nu} - K^{\nu}{}_{\rho\nu}A^{\mu\rho}\\
&= \nabla_{\nu}A^{\mu\nu} - \frac{1}{2}\left[(T_{\rho}{}^{\mu}{}_{\nu} + T_{\nu}{}^{\mu}{}_{\rho} - T^{\mu}{}_{\rho\nu})A^{\rho\nu} - (T_{\rho}{}^{\nu}{}_{\nu} + T_{\nu}{}^{\nu}{}_{\rho} - T^{\nu}{}_{\rho\nu})A^{\mu\rho}\right]\\
&= \nabla_{\nu}A^{\mu\nu} - T_{\rho}{}^{\mu}{}_{\nu}A^{\rho\nu} - T^{\nu}{}_{\nu\rho}A^{\mu\rho}\,,
\end{split}
\end{equation}
where we have used the symmetry \(A^{[\mu\nu]} = 0\) to obtain the last line. Now combining the geometric identity~\eqref{eq:metgravcons}, the divergence~\eqref{eq:divmetfield} of the gravitational field equation and the result~\eqref{eq:covdevmetfield}, one finally arrives at
\begin{multline}\label{eq:metenmomhypcons}
\lc{\nabla}_{\nu}\Theta_{\mu}{}^{\nu} + T^{\sigma}{}_{\mu\nu}(\nabla_{\rho}H_{\sigma}{}^{\nu\rho} - T^{\tau}{}_{\tau\rho}H_{\sigma}{}^{\nu\rho})\\
- \nabla_{\nu}(\nabla_{\rho}H_{\mu}{}^{\nu\rho} - T^{\tau}{}_{\tau\rho}H_{\mu}{}^{\nu\rho}) + T^{\sigma}{}_{\sigma\nu}(\nabla_{\rho}H_{\mu}{}^{\nu\rho} - T^{\tau}{}_{\tau\rho}H_{\mu}{}^{\nu\rho}) = 0\,,
\end{multline}
which agrees with~\eqref{eq:enmomhypcons} in the case of vanishing nonmetricity.

\subsection{Spacetime symmetries and cosmology}\label{ssec:symmetry}
In the previous section we have made use of the transformation laws~\eqref{eq:mettrans} of the metric and~\eqref{eq:conntrans} of the connection under infinitesimal diffeomorphisms generated by a vector field \(X^{\mu}\). The same transformation laws also find application in the discussion of symmetric spacetimes, i.e., teleparallel geometries, which are invariant under the action of particular vector fields, \(\delta_Xg_{\mu\nu} = 0\) and \(\delta_X\Gamma^{\mu}{}_{\nu\rho} = 0\)~\cite{Hohmann:2015pva,Hohmann:2019nat}. The choice of these vector fields depends on the physical situation under consideration. A few common examples can be expressed most conveniently in spherical coordinates \((t, r, \varphi, \vartheta)\): a \emph{stationary} spacetime is invariant under the (timelike) vector field \(\partial_t\); \emph{spherical} symmetry is conveyed by the three rotation generators
\begin{equation}
\sin\varphi\partial_{\vartheta} + \frac{\cos\varphi}{\tan\vartheta}\partial_{\varphi}\,, \quad
-\cos\varphi\partial_{\vartheta} + \frac{\sin\varphi}{\tan\vartheta}\partial_{\varphi}\,, \quad
-\partial_{\varphi}\,;
\end{equation}
finally, \emph{cosmological} symmetry comprises of invariance under both rotations as given above and translations, defined by the vector fields
\begin{subequations}
\begin{align}
& \chi\sin\vartheta\cos\varphi\partial_r + \frac{\chi}{r}\cos\vartheta\cos\varphi\partial_{\vartheta} - \frac{\chi\sin\varphi}{r\sin\vartheta}\partial_{\varphi}\,,\\
& \chi\sin\vartheta\sin\varphi\partial_r + \frac{\chi}{r}\cos\vartheta\sin\varphi\partial_{\vartheta} + \frac{\chi\cos\varphi}{r\sin\vartheta}\partial_{\varphi}\,,\\
& \chi\cos\vartheta\partial_r - \frac{\chi}{r}\sin\vartheta\partial_{\vartheta}\,,
\end{align}
\end{subequations}
where we used the abbreviation \(\chi = \sqrt{1 - (ur)^2}\), and \(u\) can be any real or imaginary number, so that the sign of \(u^2 \in \mathbb{R}\) determines the curvature of the spatial hypersurfaces of constant time \(t\). For \(u^2 > 0\), their spatial curvature is positive, while \(u^2 < 0\) corresponds to negative spatial curvature. Finally, \(u^2 = 0\) is the spatially flat case.

Symmetric spacetimes are often considered as potential solutions to the field equations of a given theory, since they are completely characterized by fewer functions than there are components of the dynamical fields, and these functions depend on a smaller number of coordinates, hence leading to a simple ansatz for solving the field equations. As a simple and physically well motivated example, we show this for the case of cosmological symmetry in the teleparallel geometry. It is well known that the most general metric which is homogeneous and isotropic is the Friedmann-Lemaître-Robertson-Walker metric
\begin{equation}
g_{\mu\nu} = -n_{\mu}n_{\nu} + h_{\mu\nu}\,,
\end{equation}
where the hypersurface conormal
\begin{equation}
n_{\mu}\dd x^{\mu} = -N\dd t
\end{equation}
and spatial metric
\begin{equation}
h_{\mu\nu}\dd x^{\mu} \otimes \dd x^{\nu} = A^2\left[\frac{\dd r \otimes \dd r}{\chi^2} + r^2(\dd\vartheta \otimes \dd\vartheta + \sin^2\vartheta\dd\varphi \otimes \dd\varphi)\right]
\end{equation}
are fully determined by two functions of time, known as the lapse function \(N = N(t)\) and scale factor \(A = A(t)\). Using this metric, we can apply the decomposition~\eqref{eq:conndec} of the affine connection, and we find that the most general homogeneous and isotropic connection is characterized through its torsion and nonmetricity
\begin{subequations}
\begin{align}
T^{\mu}{}_{\nu\rho} &= \frac{2}{A}(\mathcal{T}_1h^{\mu}_{[\nu}n_{\rho]} + \mathcal{T}_2n_{\sigma}\varepsilon^{\sigma\mu}{}_{\nu\rho})\,,\\
Q_{\rho\mu\nu} &= \frac{2}{A}(\mathcal{Q}_1n_{\rho}n_{\mu}n_{\nu} + 2\mathcal{Q}_2n_{\rho}h_{\mu\nu} + 2\mathcal{Q}_3h_{\rho(\mu}n_{\nu)})\,,
\end{align}
\end{subequations}
by five further functions \(\mathcal{T}_1, \mathcal{T}_2, \mathcal{Q}_1, \mathcal{Q}_2, \mathcal{Q}_3\) of time, and \(\varepsilon_{\mu\nu\rho\sigma}\) is the totally antisymmetric tensor normalized such that
\begin{equation}
\varepsilon_{0123} = \sqrt{-g} = \frac{NA^3r^2\sin\vartheta}{\chi}\,.
\end{equation}
Note that in general the curvature of this connection does not vanish, and so one must impose additional constraints on the aforementioned functions. Before discussing these constraints, it is most convenient to introduce the conformal time derivative
\begin{equation}
F' = \frac{A}{N}\frac{\dd F}{\dd t}
\end{equation}
acting on any time-dependent scalar function \(F = F(t)\), as well as the conformal Hubble parameter
\begin{equation}
\mathcal{H} = \frac{A'}{A} = \frac{1}{N}\frac{\dd A}{\dd t}\,.
\end{equation}
With the help of these definitions, the conditions on the parameter functions under which the curvature tensor vanishes become
\begin{subequations}
\begin{align}
\mathcal{T}_2(\mathcal{H} - \mathcal{T}_1 + \mathcal{Q}_2) &= 0\,,\label{eq:cosmocurv1}\\
\mathcal{T}_2(\mathcal{H} - \mathcal{T}_1 + \mathcal{Q}_2 - \mathcal{Q}_3) &= 0\,,\label{eq:cosmocurv2}\\
(\mathcal{H} - \mathcal{T}_1 + \mathcal{Q}_2)(\mathcal{H} - \mathcal{T}_1 + \mathcal{Q}_2 - \mathcal{Q}_3) - \mathcal{T}_2^2 + u^2 &= 0\,,\label{eq:cosmocurv3}\\
(\mathcal{Q}_1 + \mathcal{Q}_2)(\mathcal{H} - \mathcal{T}_1 + \mathcal{Q}_2) + (\mathcal{H} - \mathcal{T}_1 + \mathcal{Q}_2)' &= 0\,,\label{eq:cosmocurv4}\\
(\mathcal{Q}_1 + \mathcal{Q}_2)(\mathcal{H} - \mathcal{T}_1 + \mathcal{Q}_2 - \mathcal{Q}_3) - (\mathcal{H} - \mathcal{T}_1 + \mathcal{Q}_2 - \mathcal{Q}_3)' &= 0\,,\label{eq:cosmocurv5}\\
\mathcal{T}_2' &= 0\,.\label{eq:cosmocurv6}
\end{align}
\end{subequations}
Note that \(u\) appears in only one of these equations; nevertheless, it plays an important role for the solutions of this system, as we will show now. For this purpose, consider first the case \(u^2 \neq 0\). In this case the condition~\eqref{eq:cosmocurv3} implies
\begin{equation}
(\mathcal{H} - \mathcal{T}_1 + \mathcal{Q}_2)(\mathcal{H} - \mathcal{T}_1 + \mathcal{Q}_2 - \mathcal{Q}_3) \neq \mathcal{T}_2^2\,.
\end{equation}
From the two conditions~\eqref{eq:cosmocurv1} and~\eqref{eq:cosmocurv2} then further follows that either the right hand side vanishes, or both factors on the left hand side vanish. We first consider this latter case. The condition~\eqref{eq:cosmocurv3} then requires \(\mathcal{T}_2 = \pm u\), and we find that all remaining equations are solved by
\begin{equation}\label{eq:cosmogaxi}
\mathcal{T}_2 = \pm u\,, \quad
\mathcal{T}_1 - \mathcal{Q}_2 = \mathcal{H}\,, \quad
\mathcal{Q}_3 = 0\,.
\end{equation}
Also we see that we must demand \(u\) to be real in order to obtain a real value of the connection coefficients; hence, this solution is valid only for positive spatial curvature \(u^2 > 0\). Alternatively, the first two equations~\eqref{eq:cosmocurv1} and~\eqref{eq:cosmocurv2} can also be solved by setting \(\mathcal{T}_2 = 0\). From the remaining equations then follows the solution
\begin{equation}\label{eq:cosmogvec}
\mathcal{T}_2 = 0\,, \quad
(\mathcal{H} - \mathcal{T}_1 + \mathcal{Q}_2)(\mathcal{H} - \mathcal{T}_1 + \mathcal{Q}_2 - \mathcal{Q}_3) = -u^2\,, \quad
\mathcal{Q}_1 + \mathcal{Q}_2 = -\frac{\mathcal{H}' - \mathcal{T}_1' + \mathcal{Q}_2'}{\mathcal{H} - \mathcal{T}_1 + \mathcal{Q}_2}\,.
\end{equation}
Now we see that both signs of \(u^2\) are allowed. These are the only two possibilities to solve the first three equations, and so we may turn our attention to the case \(u = 0\). In this case the third equation~\eqref{eq:cosmocurv3} mandates
\begin{equation}
(\mathcal{H} - \mathcal{T}_1 + \mathcal{Q}_2)(\mathcal{H} - \mathcal{T}_1 + \mathcal{Q}_2 - \mathcal{Q}_3) = \mathcal{T}_2^2\,,
\end{equation}
and we see that both sides of this equation must vanish in order to satisfy the conditions~\eqref{eq:cosmocurv1} and~\eqref{eq:cosmocurv2}, so that all solutions will have \(\mathcal{T}_2 = 0\). For the left hand side, we are free to choose at most one of the two factors to be non-vanishing. This leads to the three possible solutions
\begin{equation}\label{eq:cosmogcoin}
\mathcal{T}_2 = 0\,, \quad
\mathcal{T}_1 - \mathcal{Q}_2 = \mathcal{H}\,, \quad
\mathcal{Q}_3 = 0
\end{equation}
if both factors vanish,
\begin{equation}\label{eq:cosmogconf}
\mathcal{T}_2 = 0\,, \quad
\mathcal{T}_1 - \mathcal{Q}_2 + \mathcal{Q}_3 = \mathcal{H}\,, \quad
\mathcal{Q}_1 + \mathcal{Q}_2 = -\frac{\mathcal{Q}_3'}{\mathcal{Q}_3}
\end{equation}
if only the second factor vanishes, as well as
\begin{equation}\label{eq:cosmogpara}
\mathcal{T}_2 = 0\,, \quad
\mathcal{T}_1 - \mathcal{Q}_2 = \mathcal{H}\,, \quad
\mathcal{Q}_1 + \mathcal{Q}_2 = \frac{\mathcal{Q}_3'}{\mathcal{Q}_3}
\end{equation}
if only the first factor vanishes. These are the only possible homogeneous and isotropic teleparallel geometries. Note that for each solution one has three conditions on the five parameter functions, so that two of them can be freely chosen to parametrize the solution, and must be determined alongside the scale factor \(A\) and lapse \(N\) by solving the field equations of a given teleparallel gravity theory.

In the discussion above we have assumed a general teleparallel geometry, for which both torsion and nonmetricity are allowed to be non-vanishing. From the solutions we have found one can now easily deduce the symmetric and metric teleparallel geometries. We start with the former, by imposing the additional condition \(\mathcal{T}_1 = \mathcal{T}_2 = 0\). One immediately sees that this condition is not compatible with the first solution~\eqref{eq:cosmogaxi}, which explicitly demands \(\mathcal{T}_2 \neq 0\), and so this solution cannot be restricted to symmetric teleparallel gravity. This is different for the remaining solutions. From the solution~\eqref{eq:cosmogvec}, one obtains the spatially curved case
\begin{equation}\label{eq:cosmosvec}
(\mathcal{H} + \mathcal{Q}_2)(\mathcal{H} + \mathcal{Q}_2 - \mathcal{Q}_3) = - u^2\,, \quad
\mathcal{Q}_1 + \mathcal{Q}_2 = -\frac{\mathcal{H}' + \mathcal{Q}_2'}{\mathcal{H} + \mathcal{Q}_2}\,,
\end{equation}
while the three spatially flat solutions become
\begin{subequations}
\begin{align}
\mathcal{Q}_2 &= -\mathcal{H}\,, &
\mathcal{Q}_3 &= 0\,, \\
\mathcal{Q}_2 - \mathcal{Q}_3 &= -\mathcal{H}\,, &
\mathcal{Q}_1 + \mathcal{Q}_2 &= -\frac{\mathcal{Q}_3'}{\mathcal{Q}_3}\,,\\
\mathcal{Q}_2 &= -\mathcal{H}\,, &
\mathcal{Q}_1 + \mathcal{Q}_2 &= \frac{\mathcal{Q}_3'}{\mathcal{Q}_3}\,.
\end{align}
\end{subequations}
For each of these solutions one has two conditions on the three scalar functions \(\mathcal{Q}_{1,2,3}\) which parametrize the nonmetricity, so that one of them remains undetermined by the symmetry condition, and is left to be determined by the gravitational field equations~\cite{DAmbrosio:2021pnd,Hohmann:2021ast}.

In a similar fashion, one can also restrict the general teleparallel cosmologies to the metric teleparallel geometry, by imposing the conditions \(\mathcal{Q}_1 = \mathcal{Q}_2 = \mathcal{Q}_3 = 0\) of vanishing nonmetricity. For the first solution~\eqref{eq:cosmogaxi}, this leads to
\begin{equation}\label{eq:cosmomaxi}
\mathcal{T}_2 = \pm u\,, \quad
\mathcal{T}_1 = \mathcal{H}\,,
\end{equation}
while the second solution~\eqref{eq:cosmogvec} becomes
\begin{equation}\label{eq:cosmomvec}
\mathcal{T}_2 = 0\,, \quad
\mathcal{T}_1 = \mathcal{H} \pm iu\,.
\end{equation}
For the latter, we have explicitly solved the appearing quadratic equation. In this case we see that \(u\) must be imaginary in order to obtain a real torsion, and so we are restricted to the case \(u^2  < 0\) of negative spatial curvature. Finally, the three solutions for \(u = 0\) reduce to the common case
\begin{equation}\label{eq:cosmomflat}
\mathcal{T}_2 = 0\,, \quad
\mathcal{T}_1 = \mathcal{H}\,.
\end{equation}
In all three cases, the two free functions in the torsion scalar are fixed by the conditions of cosmological symmetry and vanishing curvature, so that the field equations are fully expressed in terms of the scale factor \(A\) and the lapse \(N\)~\cite{Hohmann:2020zre}.

\subsection{Perturbation theory}\label{ssec:pert}
Besides the use of symmetric spacetimes as shown in the previous section, another common approach to simplify the (in general non-linear) field equations of a given teleparallel gravity theory is to start from a known, usually highly symmetric solution of the field equations, given by a metric \(\bar{g}_{\mu\nu}\) and a flat affine connection with coefficients \(\bar{\Gamma}^{\mu}{}_{\nu\rho}\), and perform a perturbative expansion of the dynamical fields and their governing field equations around this solution. For this purpose, one conventionally introduces a perturbation parameter \(\epsilon\) on which the solution will depend, and which can be related, for example, to the gravitational constant for a weak-field approximation, or the inverse speed of light for a low-velocity approximation. The full solution \(g_{\mu\nu}(\epsilon)\) and \(\Gamma^{\mu}{}_{\nu\rho}(\epsilon)\), is then expanded in a Taylor series
\begin{equation}
g_{\mu\nu} = \sum_{k = 0}^{\infty}\frac{\epsilon^k}{k!}\left.\frac{\dd^k}{\dd\epsilon^k}g_{\mu\nu}\right|_{\epsilon = 0}\,, \quad
\Gamma^{\mu}{}_{\nu\rho} = \sum_{k = 0}^{\infty}\frac{\epsilon^k}{k!}\left.\frac{\dd^k}{\dd\epsilon^k}\Gamma^{\mu}{}_{\nu\rho}\right|_{\epsilon = 0}
\end{equation}
around the background solution \(\bar{g}_{\mu\nu} = g_{\mu\nu}(0)\) and \(\bar{\Gamma}^{\mu}{}_{\nu\rho} = \Gamma^{\mu}{}_{\nu\rho}(0)\). Different conventions are abundant for the terms in this Taylor expansion, either for the coefficients
\begin{equation}
\delta^kg_{\mu\nu} = \left.\frac{\dd^k}{\dd\epsilon^k}g_{\mu\nu}\right|_{\epsilon = 0}\,, \quad
\delta^k\Gamma^{\mu}{}_{\nu\rho} = \left.\frac{\dd^k}{\dd\epsilon^k}\Gamma^{\mu}{}_{\nu\rho}\right|_{\epsilon = 0}\,,
\end{equation}
or for the full terms
\begin{equation}
\order{g}{k}_{\mu\nu} = \frac{\epsilon^k}{k!}\left.\frac{\dd^k}{\dd\epsilon^k}g_{\mu\nu}\right|_{\epsilon = 0}\,, \quad
\order{\Gamma}{k}^{\mu}{}_{\nu\rho} = \frac{\epsilon^k}{k!}\left.\frac{\dd^k}{\dd\epsilon^k}\Gamma^{\mu}{}_{\nu\rho}\right|_{\epsilon = 0}\,.
\end{equation}
In the following, we will make use of the latter, as it turns out to be shorter for the examples we consider. It follows from the fact that the metric \(g_{\mu\nu}\) is a symmetric tensor field that the same property holds also for all terms \(\order{g}{k}_{\mu\nu}\) in its perturbative expansion. For the connection coefficients, one similarly concludes from the fact that both \(\Gamma^{\mu}{}_{\nu\rho}\) and \(\bar{\Gamma}^{\mu}{}_{\nu\rho}\) are connection coefficients that the remaining terms \(\order{\Gamma}{k}^{\mu}{}_{\nu\rho}\) with \(k > 0\) are tensor fields. In order to determine these terms, one performs a similar Taylor expansion of the gravitational field equations in the perturbation parameter \(\epsilon\). It is the main virtue of this expansion that at each perturbation order \(k\), the corresponding terms of the field equations comprise a linear equation for the field terms \(\order{g}{k}_{\mu\nu}\) and \(\order{\Gamma}{k}^{\mu}{}_{\nu\rho}\) at the same order, which contain the lower order terms as a source; hence, they can be solved subsequently for increasing orders, where the previously found solutions for lower orders are used in each further order to be solved.

In the case of teleparallel gravity, it is important to keep in mind that next to the gravitational field equations also the constraint~\eqref{eq:curvature} of vanishing curvature must be satisfied at any perturbation order. In the symmetric and metric teleparallel classes of theories, also either the torsion~\eqref{eq:torsion} or nonmetricity~\eqref{eq:nonmetricity} must vanish at each order. While it is possible to simply consider these constraints as additional equations which must be solved next to the field equations at each order, one may also pose the question whether it is possible to find a general perturbative solution to these constraints, which is independent of the gravity theory under consideration, and which can then be inserted into the perturbed field equations of any specific gravity theory. To obtain this solution, one needs to perform a perturbative expansion of the corresponding constraint equations. We start by showing this procedure for the flatness constraint~\eqref{eq:curvature}. At the zeroth order, this simply becomes the vanishing of the curvature
\begin{equation}
0 = \bar{R}^{\mu}{}_{\nu\rho\sigma} = \partial_{\rho}\bar{\Gamma}^{\mu}{}_{\nu\sigma} - \partial_{\sigma}\bar{\Gamma}^{\mu}{}_{\nu\rho} + \bar{\Gamma}^{\mu}{}_{\tau\rho}\bar{\Gamma}^{\tau}{}_{\nu\sigma} - \bar{\Gamma}^{\mu}{}_{\tau\sigma}\bar{\Gamma}^{\tau}{}_{\nu\rho}
\end{equation}
for the background connection \(\bar{\Gamma}^{\mu}{}_{\nu\rho}\), which we assume to be satisfied from now on. For the first-order perturbation of the curvature, one finds the condition
\begin{equation}
0 = \order{R}{1}^{\mu}{}_{\nu\rho\sigma} = \bar{\nabla}_{\rho}\order{\Gamma}{1}^{\mu}{}_{\nu\sigma} - \bar{\nabla}_{\sigma}\order{\Gamma}{1}^{\mu}{}_{\nu\rho} + \bar{T}^{\tau}{}_{\rho\sigma}\order{\Gamma}{1}^{\mu}{}_{\nu\tau}\,,
\end{equation}
where all quantities which are calculated with respect to the background connection are denoted with a bar. Now it is helpful to recall that the commutator of covariant derivatives is given by
\begin{equation}
\bar{\nabla}_{\rho}\bar{\nabla}_{\sigma}\lambda^{\mu}{}_{\nu} - \bar{\nabla}_{\sigma}\bar{\nabla}_{\rho}\lambda^{\mu}{}_{\nu} = \bar{R}^{\mu}{}_{\tau\rho\nu}\lambda^{\tau}{}_{\nu} - \bar{R}^{\tau}{}_{\nu\rho\nu}\lambda^{\mu}{}_{\tau} - \bar{T}^{\tau}{}_{\rho\sigma}\bar{\nabla}_{\tau}\lambda^{\mu}{}_{\nu}
\end{equation}
for a tensor field \(\lambda^{\mu}{}_{\nu}\), where the two curvature terms on the right hand side vanish. Hence, we can solve the flatness condition at the first perturbation order by setting
\begin{equation}\label{eq:genpert1}
\order{\Gamma}{1}^{\mu}{}_{\nu\rho} = \bar{\nabla}_{\rho}\order{\lambda}{1}^{\mu}{}_{\nu}
\end{equation}
with an arbitrary first-order tensor field \(\order{\lambda}{1}^{\mu}{}_{\nu}\). To illustrate the further procedure, we calculate the second order curvature perturbation
\begin{equation}
0 = \order{R}{2}^{\mu}{}_{\nu\rho\sigma} = \bar{\nabla}_{\rho}\order{\Gamma}{2}^{\mu}{}_{\nu\sigma} - \bar{\nabla}_{\sigma}\order{\Gamma}{2}^{\mu}{}_{\nu\rho} + \bar{T}^{\tau}{}_{\rho\sigma}\order{\Gamma}{2}^{\mu}{}_{\nu\tau} + 2\order{\Gamma}{1}^{\mu}{}_{\tau[\rho}\order{\Gamma}{1}^{\tau}{}_{|\nu|\sigma]}\,,
\end{equation}
where we now also need to take into account the first-order connection perturbation. A naive ansatz \(\bar{\nabla}_{\rho}\order{\lambda}{2}^{\mu}{}_{\nu}\) for \(\order{\Gamma}{2}^{\mu}{}_{\nu\rho}\) is therefore not sufficient. In order to cancel the term arising from the first-order connection perturbation, one also needs to include terms which are quadratic in \(\order{\lambda}{1}^{\mu}{}_{\nu}\), and must contain one derivative. One finds that a possible solution is given by
\begin{equation}\label{eq:genpert2}
\order{\Gamma}{2}^{\mu}{}_{\nu\rho} = \bar{\nabla}_{\rho}\order{\lambda}{2}^{\mu}{}_{\nu} - \order{\lambda}{1}^{\mu}{}_{\tau}\bar{\nabla}_{\rho}\order{\lambda}{1}^{\tau}{}_{\nu}\,.
\end{equation}
Note that this solution is not unique. Alternatively, one could choose, for example,
\begin{equation}
\order{\Gamma}{2}^{\mu}{}_{\nu\rho} = \bar{\nabla}_{\rho}\order{\lambda}{2}^{\mu}{}_{\nu} + \bar{\nabla}_{\rho}\order{\lambda}{1}^{\mu}{}_{\tau}\order{\lambda}{1}^{\tau}{}_{\nu}\,.
\end{equation}
This becomes clear by realizing that these solutions differ only by a term \(\bar{\nabla}_{\rho}(\order{\lambda}{1}^{\mu}{}_{\tau}\order{\lambda}{1}^{\tau}{}_{\nu})\), which can be absorbed by a redefinition of \(\order{\lambda}{2}^{\mu}{}_{\nu}\). By a similar procedure, one can also subsequently solve the flatness condition at any higher perturbation order.

For the symmetric and metric teleparallel cases, one can make use of the already determined solution of the perturbative flatness condition, and further restrict the perturbation tensor fields \(\order{\lambda}{k}^{\mu}{}_{\nu}\) in order to achieve a connection with vanishing torsion or nonmetricity at any perturbation. We start with the former, which means that the background connection coefficients \(\bar{\Gamma}^{\mu}{}_{\nu\rho}\) as well as the perturbations \(\order{\Gamma}{k}^{\mu}{}_{\nu\rho}\) must be symmetric in their lower two indices. To obtain this property, one can make use of the result that the flat connection perturbations can always be parametrized in the form
\begin{equation}
\order{\Gamma}{k}^{\mu}{}_{\nu\rho} = \bar{\nabla}_{\rho}\order{\lambda}{k}^{\mu}{}_{\nu} + \sum_{j = 1}^{k - 1}\order{\Lambda}{j,k}^{\mu}{}_{\tau}\order{\Gamma}{j}^{\tau}{}_{\nu\rho}\,,
\end{equation}
where \(\order{\Lambda}{j,k}^{\mu}{}_{\tau}\) is determined by solving the flatness condition at the $k$'th perturbation order. Indeed, we have seen this form explicitly for the first order~\eqref{eq:genpert1}, as well as the second order~\eqref{eq:genpert2}, where for the latter \(\order{\Lambda}{1,2}^{\mu}{}_{\tau} = -\order{\lambda}{1}^{\mu}{}_{\tau}\). Using this parametrization, it follows that once we have solved the condition of vanishing torsion up to the perturbation order \(k - 1\), the condition for the $k$'th order simply becomes
\begin{equation}
\bar{\nabla}_{[\rho}\order{\lambda}{k}^{\mu}{}_{\nu]} = 0\,.
\end{equation}
Further using the fact that the covariant derivatives with respect to the background connection commute in the absence of curvature and torsion, we can thus write the solution as
\begin{equation}\label{eq:sympert}
\order{\lambda}{k}^{\mu}{}_{\nu} = \bar{\nabla}_{\nu}\order{\zeta}{k}^{\mu}\,,
\end{equation}
where we introduced the perturbation parameters \(\order{\zeta}{k}^{\mu}\). Note, however, that also in this case numerous other parametrizations of the connection coefficients can be found. A parametrization which turns out particularly convenient for practical calculations arises from the fact that two flat, torsion-free connections, are locally related by a diffeomorphism. It follows that also the perturbed connection \(\Gamma^{\mu}{}_{\nu\rho}\), which depends on the perturbation parameter \(\epsilon\), and the background \(\bar{\Gamma}^{\mu}{}_{\nu\rho}\), are locally related by a family \(\Phi_{\epsilon}\) of diffeomorphisms parametrized by \(\epsilon\), such that
\begin{equation}
\Gamma^{\mu}{}_{\nu\rho}(\epsilon) = \Phi_{\epsilon}^*\bar{\Gamma}^{\mu}{}_{\nu\rho}\,.
\end{equation}
Performing a Taylor expansion in order to obtain the perturbation terms \(\order{\Gamma}{k}^{\mu}{}_{\nu\rho}\), we see that on the right hand side \(\bar{\Gamma}^{\mu}{}_{\nu\rho}\) remains fixed, and we need to expand the diffeomorphism \(\Phi_{\epsilon}\) into a corresponding Taylor series. It turns out that such an expansion gives rise to a series of vector fields \(\order{\xi}{k}^{\mu}\) for \(k > 0\), in terms of which the Taylor expansion reads~\cite{Bruni:1996im,Sonego:1997np,Bruni:1999et}
\begin{equation}
\order{\Gamma}{k}^{\mu}{}_{\nu\rho} = \sum_{l_1 + 2l_2 + \ldots = k}\frac{1}{l_1!l_2! \cdots}\left(\mathcal{L}_{\order{\xi}{1}}^{l_1} \cdots \mathcal{L}_{\order{\xi}{j}}^{l_j} \cdots \bar{\Gamma}\right)^{\mu}{}_{\nu\rho}\,.
\end{equation}
It is instructive to calculate the lower order terms explicitly, using the formula~\eqref{eq:conntrans} with vanishing curvature and torsion. For the background, the expansion trivially reduces to
\begin{equation}
\order{\Gamma}{0}^{\mu}{}_{\nu\rho} = \bar{\Gamma}^{\mu}{}_{\nu\rho}\,.
\end{equation}
For the first order, only one term appears on the right-hand side, which reads
\begin{equation}\label{eq:sympert1}
\order{\Gamma}{1}^{\mu}{}_{\nu\rho} = \left(\mathcal{L}_{\order{\xi}{1}}\bar{\Gamma}\right)^{\mu}{}_{\nu\rho} = \bar{\nabla}_{\nu}\bar{\nabla}_{\rho}\order{\xi}{1}^{\mu}\,,
\end{equation}
while for the second order one finds the two terms
\begin{equation}
\begin{split}\label{eq:sympert2}
\order{\Gamma}{2}^{\mu}{}_{\nu\rho} &= \left(\mathcal{L}_{\order{\xi}{2}}\bar{\Gamma}\right)^{\mu}{}_{\nu\rho} + \frac{1}{2}\left(\mathcal{L}_{\order{\xi}{1}}\mathcal{L}_{\order{\xi}{1}}\bar{\Gamma}\right)^{\mu}{}_{\nu\rho}\\
&= \bar{\nabla}_{\nu}\bar{\nabla}_{\rho}\order{\xi}{2}^{\mu} + \bar{\nabla}_{(\nu}\order{\xi}{1}^{\sigma}\bar{\nabla}_{\rho)}\bar{\nabla}_{\sigma}\order{\xi}{1}^{\mu} - \frac{1}{2}\bar{\nabla}_{\nu}\bar{\nabla}_{\rho}\order{\xi}{1}^{\sigma}\bar{\nabla}_{\sigma}\order{\xi}{1}^{\mu} + \frac{1}{2}\order{\xi}{1}^{\sigma}\bar{\nabla}_{\nu}\bar{\nabla}_{\rho}\bar{\nabla}_{\sigma}\order{\xi}{1}^{\mu}\,.
\end{split}
\end{equation}
It is also helpful to compare these formulas with the perturbations~\eqref{eq:genpert1} and~\eqref{eq:genpert2}, together with the substitution~\eqref{eq:sympert}. For the first perturbation order, the perturbation~\eqref{eq:genpert1} becomes
\begin{equation}
\order{\Gamma}{1}^{\mu}{}_{\nu\rho} = \bar{\nabla}_{\nu}\bar{\nabla}_{\rho}\order{\zeta}{1}^{\mu}\,,
\end{equation}
which agrees with~\eqref{eq:sympert1} for \(\order{\zeta}{1}^{\mu} = \order{\xi}{1}^{\mu}\). At the second order, the perturbation~\eqref{eq:genpert2} becomes
\begin{equation}
\order{\Gamma}{2}^{\mu}{}_{\nu\rho} = \bar{\nabla}_{\nu}\bar{\nabla}_{\rho}\order{\zeta}{2}^{\mu} - \bar{\nabla}_{\sigma}\order{\zeta}{1}^{\mu}\bar{\nabla}_{\nu}\bar{\nabla}_{\rho}\order{\zeta}{1}^{\sigma}\,,
\end{equation}
which agrees with the result~\eqref{eq:sympert2} for
\begin{equation}
\order{\zeta}{2}^{\mu} = \order{\xi}{2}^{\mu} + \frac{1}{2}\order{\xi}{1}^{\sigma}\bar{\nabla}_{\sigma}\order{\xi}{1}^{\mu}\,.
\end{equation}
Also one easily checks that the curvature and torsion vanish at any perturbation order. This type of perturbative expansion is used, for example, to determine the propagation of gravitational waves~\cite{Hohmann:2018wxu} and the post-Newtonian limit~\cite{Flathmann:2020zyj,Hohmann:2021rmp}.

Finally, we take a look at the form of the perturbations in the metric teleparallel case, which means that the nonmetricity must vanish at all perturbation orders. This holds in particular for the background,
\begin{equation}
0 = \bar{Q}_{\mu\nu\rho} = \bar{\nabla}_{\mu}\bar{g}_{\nu\rho}\,,
\end{equation}
and so raising and lowering indices of the perturbations with the metric commutes with the covariant derivative, which greatly simplifies the calculations. We make use of this fact for calculating the conditions on the connection perturbation \(\order{\lambda}{k}^{\mu}{}_{\nu}\) which we need to satisfy in order to obtain vanishing nonmetricity. At the linear order, the perturbation of the nonmetricity is given by
\begin{equation}
\order{Q}{1}_{\mu\nu\rho} = \bar{\nabla}_{\mu}\left(\order{g}{1}_{\nu\rho} - 2\order{\lambda}{1}_{(\nu\rho)}\right)\,,
\end{equation}
and so it vanishes if we fix the symmetric part of the connection perturbation by the condition
\begin{equation}
2\order{\lambda}{1}_{(\mu\nu)} = \order{g}{1}_{\mu\nu}\,.
\end{equation}
One could naively conclude that the same formula holds identically also for higher orders, replacing the first perturbation order with an arbitrary order \(k\). However, this is not the case, as follows from a perturbative expansion of the nonmetricity~\eqref{eq:nonmetricity}, whose higher than linear orders contain products of the lower order perturbations of the metric and the connection. For example, for the second order we have
\begin{equation}
\begin{split}
\order{Q}{2}_{\mu\nu\rho} &= \bar{\nabla}_{\mu}\left(\order{g}{2}_{\nu\rho} - 2\order{\lambda}{2}_{(\nu\rho)}\right) - 2\nabla_{\mu}\order{\lambda}{1}^{\sigma}{}_{(\nu}\left(\order{g}{1}_{\rho)\sigma} - \order{\lambda}{1}_{\rho)\sigma}\right)\\
&= \bar{\nabla}_{\mu}\left(\order{g}{2}_{\nu\rho} - 2\order{\lambda}{2}_{(\nu\rho)} - \bar{g}_{\sigma\omega}\order{\lambda}{1}^{\sigma}{}_{\nu}\order{\lambda}{1}^{\omega}{}_{\rho}\right)\,,
\end{split}
\end{equation}
after substituting \(\order{g}{1}_{\nu\rho}\) from the first order result, so that we can easily read off the condition for the second order perturbation. Following the same procedure also for higher order perturbations, one arrives at the general formula
\begin{equation}
2\order{\lambda}{k}_{(\mu\nu)} = \order{g}{k}_{\mu\nu} - \bar{g}_{\rho\sigma}\sum_{j = 1}^{k - 1}\order{\lambda}{j}^{\rho}{}_{\mu}\order{\lambda}{k - j}^{\sigma}{}_{\nu}\,.
\end{equation}
We see that the condition of vanishing nonmetricity links the symmetric part of the connection perturbation to the perturbation of the metric, and so the latter can always be expressed in terms of the former, leaving \(\order{\lambda}{k}_{\mu\nu}\) as the only independent perturbation variable. Like in the case of symmetric teleparallel gravity, this perturbative expansion is used for the calculation of gravitational waves~\cite{Hohmann:2018jso} and the post-Newtonian limit~\cite{Ualikhanova:2019ygl}, but also in the cosmological perturbation theory around flat~\cite{Golovnev:2018wbh} and general~\cite{Bahamonde:2022ohm} cosmological backgrounds.

\section{Teleparallel gravity theories}\label{sec:theories}
In this final section, we discuss a few selected classes of teleparallel gravity theories, their actions and field equations. These theories constitute modifications of general relativity, which depart from a reformulation of the Einstein-Hilbert action in terms of teleparallel geometries, known as the teleparallel equivalent of general relativity, which we discuss in section~\ref{ssec:tegr}. A simple modification is then obtained by replacing the Lagrangian of these theories by a free function thereof, as we show in section~\ref{ssec:fg}. Another modification arises by considering the most general action which is quadratic in torsion and nonmetricity, and which we show in section~\ref{ssec:quadlag}. Moreover, modified theories can be obtained by considering a scalar field as another dynamical variable in addition to the metric and the flat connection; we discuss theories of this type in section~\ref{ssec:scalartele}, and see how a particular subclass of them is connected to a previously discussed class of theories in section~\ref{ssec:stfg}.

\subsection{The teleparallel equivalents of general relativity}\label{ssec:tegr}
In the previous sections we have discussed the general form of the action and field equations for teleparallel gravity theories, but we have not yet considered any particular theories. As a starting point for the construction of modified teleparallel gravity theories, we now pose the question how the well-known general relativity action and field equations can be cast into the teleparallel framework. The crucial observation to answer this question is the fact that the decomposition~\eqref{eq:conndec} of the independent connection with respect to the Levi-Civita connection of the metric induces a related decomposition of the curvature given by
\begin{equation}
R^{\mu}{}_{\nu\rho\sigma} = \lc{R}^{\mu}{}_{\nu\rho\sigma} + \lc{\nabla}_{\rho}M^{\mu}{}_{\nu\sigma} - \lc{\nabla}_{\sigma}M^{\mu}{}_{\nu\rho} + M^{\mu}{}_{\tau\rho}M^{\tau}{}_{\nu\sigma} - M^{\mu}{}_{\tau\sigma}M^{\tau}{}_{\nu\rho}\,.
\end{equation}
Keeping in mind that the curvature~\eqref{eq:curvature} of the teleparallel connection is imposed to vanish, one can solve for the curvature tensor of the Levi-Civita connection, and finds
\begin{equation}\label{eq:curvdec}
\lc{R}^{\mu}{}_{\nu\rho\sigma} = -\lc{\nabla}_{\rho}M^{\mu}{}_{\nu\sigma} + \lc{\nabla}_{\sigma}M^{\mu}{}_{\nu\rho} - M^{\mu}{}_{\tau\rho}M^{\tau}{}_{\nu\sigma} + M^{\mu}{}_{\tau\sigma}M^{\tau}{}_{\nu\rho}\,.
\end{equation}
This allows us to replace the Ricci scalar \(\lc{R}\) in the Einstein-Hilbert action
\begin{equation}\label{eq:einsteinhilbert}
S_{\text{g}} = \frac{1}{2\kappa^2}\int_M\lc{R}\sqrt{-g}\dd^4x
\end{equation}
by
\begin{equation}\label{eq:riccisplit}
\lc{R} = -G + B\,,
\end{equation}
where we defined the terms
\begin{equation}\label{eq:gtegrbndterms}
G = 2M^{\mu}{}_{\tau[\mu}M^{\tau\nu}{}_{\nu]}\,, \quad
B = 2\lc{\nabla}_{\mu}M^{[\nu\mu]}{}_{\nu}\,.
\end{equation}
One can see that \(B\) becomes a boundary term in the action, and therefore does not contribute to the field equations. Omitting this term from the action, one thus obtains~\cite{BeltranJimenez:2019odq}
\begin{equation}\label{eq:gtegraction}
S_{\text{g}} = -\frac{1}{2\kappa^2}\int_MG\sqrt{-g}\dd^4x\,.
\end{equation}
This is the action of the \emph{general teleparallel equivalent of general relativity} (GTEGR). To study the nature of this equivalence, we calculate the gravitational field equations. Note that the variation of the distortion tensor is given by
\begin{equation}
\delta M^{\mu}{}_{\nu\rho} = \delta\Gamma^{\mu}{}_{\nu\rho} - \delta\lc{\Gamma}^{\mu}{}_{\nu\rho} = \delta\Gamma^{\mu}{}_{\nu\rho} - \frac{1}{2}g^{\mu\sigma}\left(\lc{\nabla}_{\nu}\delta g_{\sigma\rho} + \lc{\nabla}_{\rho}\delta g_{\nu\sigma} - \lc{\nabla}_{\sigma}\delta g_{\nu\rho}\right)\,,
\end{equation}
and so the variation of the gravity scalar becomes
\begin{equation}
\delta G = U^{\mu\nu}\delta g_{\mu\nu} + V^{\rho\mu\nu}\lc{\nabla}_{\rho}\delta g_{\mu\nu} + Z_{\mu}{}^{\nu\rho}\delta\Gamma^{\mu}{}_{\nu\rho}\,,
\end{equation}
where we have introduced the abbreviations
\begin{subequations}\label{eq:gvarabbrev}
\begin{align}
U^{\mu\nu} &= M^{\rho\sigma(\mu}M_{\sigma}{}^{\nu)}{}_{\rho} - M^{\rho(\mu\nu)}M^{\sigma}{}_{\rho\sigma}\\
V^{\rho\mu\nu} &= M^{\rho(\mu\nu)} - M^{\sigma(\mu}{}_{\sigma}g^{\nu)\rho} - M^{[\rho\sigma]}{}_{\sigma}g^{\mu\nu}\\
Z_{\mu}{}^{\nu\rho} &= M^{\nu\sigma}{}_{\sigma}\delta_{\mu}^{\rho} + M^{\sigma}{}_{\mu\sigma}g^{\nu\rho} - M^{\nu\rho}{}_{\mu} - M^{\rho}{}_{\mu}{}^{\nu}\,.
\end{align}
\end{subequations}
This allows us to calculate the variation of the action~\eqref{eq:gtegraction} and perform integration by parts in order to eliminate the derivatives acting on the metric perturbation \(\delta g_{\mu\nu}\). The resulting variation then takes the form~\eqref{eq:metricgravactvar} with
\begin{equation}\label{eq:gtegrmetvar}
\begin{split}
W_{\mu\nu} &= \frac{1}{\kappa^2}\left(U_{\mu\nu} - \lc{\nabla}_{\rho}V^{\rho}{}_{\mu\nu} + \frac{1}{2}Gg_{\mu\nu}\right)\\
&= \frac{1}{\kappa^2}\bigg[\lc{\nabla}_{(\mu}M^{\rho}{}_{\nu)\rho} - \lc{\nabla}_{\rho}M^{\rho}{}_{(\mu\nu)} + M^{\rho}{}_{\sigma(\mu}M^{\sigma}{}_{\nu)\rho} - M^{\rho}{}_{\sigma\rho}M^{\sigma}{}_{(\mu\nu)}\\
&\phantom{=}- \frac{1}{2}\left(\lc{\nabla}_{\rho}M^{\sigma\rho}{}_{\sigma} - \lc{\nabla}_{\rho}M^{\rho\sigma}{}_{\sigma} + M^{\rho\sigma\omega}M_{\omega\rho\sigma} - M^{\rho}{}_{\omega\rho}M^{\omega\sigma}{}_{\sigma}\right)g_{\mu\nu}\bigg]
\end{split}
\end{equation}
and
\begin{equation}\label{eq:gtegrconnvar}
Y_{\mu}{}^{\nu\rho} = \frac{1}{2\kappa^2}Z_{\mu}{}^{\nu\rho} = \frac{1}{2\kappa^2}(M^{\nu\sigma}{}_{\sigma}\delta_{\mu}^{\rho} + M^{\sigma}{}_{\mu\sigma}g^{\nu\rho} - M^{\nu\rho}{}_{\mu} - M^{\rho}{}_{\mu}{}^{\nu})\,.
\end{equation}
By comparing with the relation~\eqref{eq:curvdec}, one finds that the first equation can be rewritten as
\begin{equation}
W_{\mu\nu} = \frac{1}{\kappa^2}\left(\lc{R}_{\mu\nu} - \frac{1}{2}\lc{R}g_{\mu\nu}\right)\,,
\end{equation}
and so the metric equation resembles Einstein's equation
\begin{equation}\label{eq:einstein}
\lc{R}_{\mu\nu} - \frac{1}{2}\lc{R}g_{\mu\nu} = \kappa^2\Theta_{\mu\nu}\,.
\end{equation}
We also need to consider the connection field equation~\eqref{eq:genconnfield}, which becomes
\begin{multline}
\frac{1}{\kappa^2}\left(\lc{\nabla}_{[\mu}M^{\nu\rho}{}_{\rho]} + \lc{\nabla}^{[\nu}M_{\rho\mu}{}^{\rho]} + M^{\nu}{}_{\rho[\mu}M^{\rho\sigma}{}_{\sigma]} + M_{\rho}{}^{\sigma[\nu}M_{\sigma\mu}{}^{\rho]}\right)\\
= \nabla_{\tau}H_{\mu}{}^{\nu\tau} - M^{\omega}{}_{\tau\omega}H_{\mu}{}^{\nu\tau}\,.
\end{multline}
Once again making use of the relation~\eqref{eq:curvdec}, the term in brackets becomes
\begin{equation}
\lc{R}^{\nu\rho}{}_{\mu\rho} + \lc{R}_{\rho\mu}{}^{\nu\rho} = \lc{R}^{\nu}{}_{\mu} - \lc{R}_{\mu}{}^{\nu} = 0\,,
\end{equation}
which vanishes, since the Ricci tensor of the Levi-Civita connection is symmetric. Hence, one is left with the equation
\begin{equation}\label{eq:gtegrhypmom}
\nabla_{\tau}H_{\mu}{}^{\nu\tau} - M^{\omega}{}_{\tau\omega}H_{\mu}{}^{\nu\tau} = 0
\end{equation}
for the hypermomentum, which must be satisfied for any matter which is compatible with the gravitational action~\eqref{eq:gtegraction}. Note that the connection does not appear anywhere on the gravitational side of the field equations, due to the fact that it enters into the action only through a total derivative term. For consistency, one conventionally assumes that it does not couple to the matter fields, so that the hypermomentum vanishes, and so the constraint~\eqref{eq:gtegrhypmom} is satisfied identically. The only non-trivial field equation is then Einstein's equation~\eqref{eq:einstein}, and so the field equations of GTEGR are equivalent to those of general relativity, hence justifying the name teleparallel equivalent.

Since the connection only has a spurious appearance in the action~\eqref{eq:gtegraction}, one may expect that it will not enter the field equations also in the symmetric and metric classes of teleparallel gravity theories. This is not obvious from the Lagrange multiplier approach of deriving the field equations, since the Lagrange multiplier terms~\eqref{eq:lagmultors} and~\eqref{eq:lagmulnonmet} are not total derivatives, and so the connection enters the field equations obtained by variation with respect to the Lagrange multipliers. Nevertheless, keeping in mind that this approach yields the same field equations as the approach of restricted variation, and that in the latter the variation of the connection appears only through a total derivative in the action, one may still expect to obtain an equivalent of general relativity. This is particularly easy to see in the case of symmetric teleparallel gravity, since its metric field equation~\eqref{eq:genmetfield} takes the same form as in the case of general teleparallel gravity, and hence once again resembles the Einstein equations~\eqref{eq:einstein}, irrespective of the constraint \(T^{\mu}{}_{\nu\rho} = 0\) imposed on the connection. For the remaining field equation~\eqref{eq:symconnfield}, it is helpful to recall that for the variation~\eqref{eq:gtegrconnvar} the left hand side of the field equation~\eqref{eq:genconnfield} vanishes identically, and hence does the left hand side of the equivalent field equation~\eqref{eq:genconnfielddens}. In the absence of torsion, the torsion term vanishes, and one is left with
\begin{equation}
\nabla_{\rho}\tilde{Y}_{\mu}{}^{\nu\rho} = 0\,.
\end{equation}
Hence, also the left hand side of the symmetric teleparallel field equation~\eqref{eq:symconnfield} vanishes identically, leaving only the hypermomentum constraint
\begin{equation}
\nabla_{\nu}\nabla_{\rho}\tilde{H}_{\mu}{}^{\nu\rho} = 0\,,
\end{equation}
which can be satisfied by demanding vanishing hypermomentum.

A similar argument holds in the case of metric teleparallel gravity. Once again, one can make use of the fact that the left hand side of the field equation~\eqref{eq:genconnfielddens} vanishes identically for the variation~\eqref{eq:gtegrconnvar}. In the absence of nonmetricity, the covariant derivative~\eqref{eq:covderdens} of the density factor \(\sqrt{-g}\) vanishes, and so this factor can be canceled from the equations. One is then left with the equation
\begin{equation}
\nabla_{\tau}Y_{\mu}{}^{\nu\tau} - T^{\omega}{}_{\omega\tau}Y_{\mu}{}^{\nu\tau} = 0\,.
\end{equation}
Using this result, the metric teleparallel field equation~\eqref{eq:metallfield} reduces to
\begin{equation}
W^{\mu\nu} = \Theta^{\mu\nu} - \nabla_{\rho}H^{\mu\nu\rho} + H^{\mu\nu\rho}T^{\tau}{}_{\tau\rho}\,.
\end{equation}
Demanding once again vanishing hypermomentum, one therefore obtains Einstein's equation~\eqref{eq:einstein} also in this case.

In order to gain more insight into the underlying structure of the different teleparallel equivalents of general relativity, it is helpful to decompose the gravity scalar \(G\) and the boundary term \(B\) into the individual contributions from the torsion and the nonmetricity. Using the connection decomposition~\eqref{eq:conndec}, the contortion~\eqref{eq:contortion} and the disformation~\eqref{eq:disformation}, one finds
\begin{multline}\label{eq:gtegrscalar}
G = \frac{1}{4}Q^{\mu\nu\rho}Q_{\mu\nu\rho} - \frac{1}{2}Q^{\mu\nu\rho}Q_{\rho\mu\nu} - \frac{1}{4}Q^{\rho\mu}{}_{\mu}Q_{\rho\nu}{}^{\nu} + \frac{1}{2}Q^{\mu}{}_{\mu\rho}Q^{\rho\nu}{}_{\nu}\\
+ \frac{1}{4}T^{\mu\nu\rho}T_{\mu\nu\rho} + \frac{1}{2}T^{\mu\nu\rho}T_{\rho\nu\mu} - T^{\mu}{}_{\mu\rho}T_{\nu}{}^{\nu\rho} + T^{\mu\nu\rho}Q_{\nu\rho\mu} - T^{\mu}{}_{\rho\mu}Q_{\rho\nu}{}^{\nu} + T^{\mu}{}_{\rho\mu}Q^{\nu}{}_{\nu\rho}\,,
\end{multline}
as well as
\begin{equation}
B = \lc{\nabla}_{\mu}(2T_{\nu}{}^{\nu\mu} + Q_{\nu}{}^{\nu\mu} - Q^{\mu\nu}{}_{\nu})\,.
\end{equation}
If either torsion or nonmetricity vanish, these expressions simplify. In particular, the gravity scalar~\eqref{eq:gtegrscalar} reduces to the nonmetricity scalar
\begin{equation}\label{eq:stegrscalar}
\begin{split}
Q &= \frac{1}{2}Q_{\rho\mu\nu}P^{\rho\mu\nu}\\
&= \frac{1}{4}Q^{\mu\nu\rho}Q_{\mu\nu\rho} - \frac{1}{2}Q^{\mu\nu\rho}Q_{\rho\mu\nu} - \frac{1}{4}Q^{\rho\mu}{}_{\mu}Q_{\rho\nu}{}^{\nu} + \frac{1}{2}Q^{\mu}{}_{\mu\rho}Q^{\rho\nu}{}_{\nu}
\end{split}
\end{equation}
or the torsion scalar
\begin{equation}\label{eq:mtegrscalar}
\begin{split}
T &= \frac{1}{2}T^{\rho}{}_{\mu\nu}S_{\rho}{}^{\mu\nu}\\
&= \frac{1}{4}T^{\mu\nu\rho}T_{\mu\nu\rho} + \frac{1}{2}T^{\mu\nu\rho}T_{\rho\nu\mu} - T^{\mu}{}_{\mu\rho}T_{\nu}{}^{\nu\rho}\,,
\end{split}
\end{equation}
respectively, where we have introduced the nonmetricity conjugate
\begin{equation}\label{eq:nonmetconj}
P^{\rho\mu\nu} = L^{\rho\mu\nu} - \frac{1}{2}g^{\mu\nu}(Q^{\rho\sigma}{}_{\sigma} - Q_{\sigma}{}^{\sigma\rho}) + \frac{1}{2}g^{\rho(\mu}Q^{\nu)\sigma}{}_{\sigma}
\end{equation}
and the superpotential
\begin{equation}\label{eq:suppot}
S_{\rho}{}^{\mu\nu} = K^{\mu\nu}{}_{\rho} - \delta_{\rho}^{\mu}T_{\sigma}{}^{\sigma\nu} + \delta_{\rho}^{\nu}T_{\sigma}{}^{\sigma\mu}\,.
\end{equation}
In terms of these scalars, the action of the \emph{symmetric teleparallel equivalent of general relativity} (STEGR) becomes~\cite{Nester:1998mp}
\begin{equation}\label{eq:stegraction}
S_{\text{g}} = -\frac{1}{2\kappa^2}\int_MQ\sqrt{-g}\dd^4x\,,
\end{equation}
while for the \emph{metric teleparallel equivalent of general relativity} (MTEGR\footnote{In the literature, the abbreviation TEGR is more common, since it was developed prior to the other equivalent theories.}) one has~\cite{Maluf:2013gaa}
\begin{equation}\label{eq:mtegraction}
S_{\text{g}} = -\frac{1}{2\kappa^2}\int_MT\sqrt{-g}\dd^4x\,.
\end{equation}
Within their respective class of teleparallel gravity theories, these actions yield the same metric field equation as general relativity, and are thus common starting points for the construction of modified gravity theories, as we will see in the following sections.

\subsection{The $f(G)$ classes of modified theories}\label{ssec:fg}
After discussing in the previous section a number of teleparallel gravity theories, whose metric field equation reproduces Einstein's field equation of general relativity for matter without hypermomentum, we now turn our focus towards modifications of these gravity theories. For the Einstein-Hilbert action~\eqref{eq:einsteinhilbert}, a well-known and thoroughly studied class of gravity theories is obtained by replacing the Ricci scalar \(\lc{R}\) by \(f(\lc{R})\), where \(f\) is an arbitrary real function of one variable, which is chosen such that the phenomenology of the resulting theory matches with observations, e.g., in cosmology. The same procedure can also be applied to the teleparallel equivalent theories~\cite{Boehmer:2021aji}. Starting with the GTEGR action~\eqref{eq:gtegraction}, one thus obtains the action
\begin{equation}\label{eq:fgaction}
S_{\text{g}} = -\frac{1}{2\kappa^2}\int_Mf(G)\sqrt{-g}\dd^4x\,.
\end{equation}
In order to derive the field equations, one proceeds as shown in the previous section, by variation of the action and integration by parts, so that the gravitational part \(S_{\text{g}}\) takes the form~\eqref{eq:metricgravactvar}, with
\begin{equation}\label{eq:fgmetvar}
W_{\mu\nu} = \frac{1}{\kappa^2}\left[f'U_{\mu\nu} - \lc{\nabla}_{\rho}(f'V^{\rho}{}_{\mu\nu}) + \frac{1}{2}fg_{\mu\nu}\right]
\end{equation}
and
\begin{equation}\label{eq:fgconnvar}
Y_{\mu}{}^{\nu\rho} = \frac{1}{2\kappa^2}f'Z_{\mu}{}^{\nu\rho}\,.
\end{equation}
where we wrote \(f, f', \ldots\) as a shorthand for \(f(G), f'(G), \ldots\), and used the abbreviations~\eqref{eq:gvarabbrev} we introduced for the variation of the gravity scalar \(G\). Hence, it follows that the gravitational field equations are given by the metric equation
\begin{equation}\label{eq:fgmetfield}
f'U_{\mu\nu} - \lc{\nabla}_{\rho}(f'V^{\rho}{}_{\mu\nu}) + \frac{1}{2}fg_{\mu\nu} = \kappa^2\Theta_{\mu\nu}
\end{equation}
and the connection equation
\begin{equation}\label{eq:fgconnfield}
\nabla_{\rho}(f'Z_{\mu}{}^{\nu\rho}) - f'M^{\omega}{}_{\rho\omega}Z_{\mu}{}^{\nu\rho} = 2\kappa^2(\nabla_{\rho}H_{\mu}{}^{\nu\rho} - M^{\omega}{}_{\rho\omega}H_{\mu}{}^{\nu\rho})\,.
\end{equation}
These equations can be written more explicitly as follows. First, recall that
\begin{equation}
U_{\mu\nu} - \lc{\nabla}_{\rho}(V^{\rho}{}_{\mu\nu}) + \frac{1}{2}Gg_{\mu\nu} = \lc{R}_{\mu\nu} - \frac{1}{2}\lc{R}g_{\mu\nu}
\end{equation}
is the left hand side of the GTEGR field equation. Using this fact, the metric field equation~\eqref{eq:fgmetfield} becomes
\begin{equation}
f'\left(\lc{R}_{\mu\nu} - \frac{1}{2}\lc{R}g_{\mu\nu}\right) - V^{\rho}{}_{\mu\nu}\lc{\nabla}_{\rho}f' + \frac{1}{2}(f - f'G)g_{\mu\nu} = \kappa^2\Theta_{\mu\nu}\,.
\end{equation}
Finally, substituting \(V^{\rho}{}_{\mu\nu}\) using the variation~\eqref{eq:gvarabbrev}, one obtains
\begin{multline}\label{eq:fgmetfield2}
f'\left(\lc{R}_{\mu\nu} - \frac{1}{2}\lc{R}g_{\mu\nu}\right) - M^{\rho}{}_{(\mu\nu)}\lc{\nabla}_{\rho}f' + \lc{\nabla}_{(\mu}f'M^{\sigma}{}_{\nu)\sigma} + M^{[\rho\sigma]}{}_{\sigma}g_{\mu\nu}\lc{\nabla}_{\rho}f'\\
+ \frac{1}{2}(f - f'G)g_{\mu\nu} = \kappa^2\Theta_{\mu\nu}\,.
\end{multline}
Similarly, one can use the fact that the left hand side of the GTEGR connection equation vanishes, and hence
\begin{equation}
\nabla_{\rho}(Z_{\mu}{}^{\nu\rho}) - M^{\omega}{}_{\rho\omega}Z_{\mu}{}^{\nu\rho} = 0\,,
\end{equation}
to write the connection field equation as
\begin{equation}
Z_{\mu}{}^{\nu\rho}\nabla_{\rho}f' = 2\kappa^2(\nabla_{\rho}H_{\mu}{}^{\nu\rho} - M^{\omega}{}_{\rho\omega}H_{\mu}{}^{\nu\rho})\,,
\end{equation}
and substituting the variation~\eqref{eq:gvarabbrev},
\begin{multline}\label{eq:fgconnfield2}
M^{\nu\rho}{}_{\rho}\nabla_{\mu}f' + M^{\sigma}{}_{\mu\sigma}g^{\nu\rho}\nabla_{\rho}f' - M^{\nu\rho}{}_{\mu}\nabla_{\rho}f' - M^{\rho}{}_{\mu}{}^{\nu}\nabla_{\rho}f'\\
= 2\kappa^2(\nabla_{\rho}H_{\mu}{}^{\nu\rho} - M^{\omega}{}_{\rho\omega}H_{\mu}{}^{\nu\rho})\,.
\end{multline}
The most important difference which distinguishes the \(f(G)\) class of theories from GTEGR is the fact that for \(f'' \neq 0\) the connection contribution to the action is no longer a total derivative, and so the connection remains as a dynamical field in the field equations, which now also contain a non-trivial connection field equation. It follows in particular that these field equations are not equivalent to those of \(f(\lc{R})\) gravity, since the latter has the metric as its only dynamical field, and its field equations are of fourth derivative order\footnote{They can be reduced to second order by introducing an auxiliary scalar field.}. In contrast, the field equations of \(f(G)\) gravity are of second derivative order.

In analogy to the GTEGR action~\eqref{eq:gtegraction}, which is based on the general teleparallel geometry containing both torsion and nonmetricity, also the teleparallel equivalent theories based on more restricted geometries can be generalized by introducing a free function into their respective actions~\eqref{eq:stegraction} and~\eqref{eq:mtegraction}. Equivalently, one can take the action~\eqref{eq:fgaction} and impose the vanishing torsion or nonmetricity either by introducing Lagrange multipliers or by imposing the constraint alongside a restricted variation. It turns out that the resulting field equations can be simplified, as we will see in the following. We start with the symmetric teleparallel gravity action~\cite{BeltranJimenez:2017tkd}
\begin{equation}\label{eq:fqaction}
S_{\text{g}} = -\frac{1}{2\kappa^2}\int_Mf(Q)\sqrt{-g}\dd^4x\,.
\end{equation}
The variation of this action is still given by the expressions~\eqref{eq:fgmetvar} and~\eqref{eq:fgconnvar}, but these simplify due to the vanishing torsion, and can be expressed in terms of the nonmetricity. Also the field equation simplify, which one can see as follows, starting with the connection equation. From the general form~\eqref{eq:symconnfield} follows that only the symmetric part \(\tilde{Y}_{\mu}{}^{(\nu\rho)}\) contributes, since the covariant derivatives commute in the absence of curvature and torsion. Using~\eqref{eq:fgconnvar}, this means that only the symmetric part \(Z_{\mu}{}^{(\nu\rho)}\) contributes, which is given by
\begin{equation}\label{eq:symconnvarrel}
Z_{\mu}{}^{(\nu\rho)} = Q_{\mu}{}^{\nu\rho} - \frac{1}{2}g^{\nu\rho}Q_{\mu\sigma}{}^{\sigma} + \frac{1}{2}\delta^{(\nu}_{\mu}Q^{\rho)\sigma}{}_{\sigma} - Q_{\sigma}{}^{\sigma(\nu}\delta^{\rho)}_{\mu} = -2P^{(\nu\rho)}{}_{\mu}\,.
\end{equation}
The connection equation therefore becomes
\begin{equation}\label{eq:fqconnfield}
-\nabla_{\nu}\nabla_{\rho}(f'\tilde{P}^{\nu\rho}{}_{\mu}) = \kappa^2\nabla_{\nu}\nabla_{\rho}\tilde{H}_{\mu}{}^{\nu\rho}\,,
\end{equation}
where \(\tilde{P}^{\nu\rho}{}_{\mu} = \sqrt{-g}P^{\nu\rho}{}_{\mu}\), and we omitted the symmetrization brackets around the indices, which are redundant due to the contraction with the commuting derivatives. Similarly, we can also simplify the metric field equation, which takes the same form~\eqref{eq:fgmetfield}, and can equivalently be written as
\begin{equation}
f'U^{\mu}{}_{\nu} - \frac{\lc{\nabla}_{\rho}(\sqrt{-g}f'V^{\rho\mu}{}_{\nu})}{\sqrt{-g}} + \frac{1}{2}f\delta^{\mu}_{\nu} = \kappa^2\Theta^{\mu}{}_{\nu}\,,
\end{equation}
using the fact that the Levi-Civita connection is metric compatible, so that we can raise and lower indices and introduce the density factor \(\sqrt{-g}\) inside the derivative. Changing this covariant derivative to the independent connection, one has
\begin{multline}
\lc{\nabla}_{\rho}(\sqrt{-g}f'V^{\rho\mu}{}_{\nu}) = \nabla_{\rho}(\sqrt{-g}f'V^{\rho\mu}{}_{\nu})\\
+ \sqrt{-g}f'(L^{\sigma}{}_{\sigma\rho}V^{\rho\mu}{}_{\nu} - L^{\rho}{}_{\sigma\rho}V^{\sigma\mu}{}_{\nu} - L^{\mu}{}_{\sigma\rho}V^{\rho\sigma}{}_{\nu} + L^{\sigma}{}_{\nu\rho}V^{\rho\mu}{}_{\sigma})\,.
\end{multline}
Calculating the variation terms
\begin{equation}
U^{\mu\nu} = \frac{1}{4}Q^{\mu\rho\sigma}Q^{\nu}{}_{\rho\sigma} + \frac{1}{4}(Q^{\rho\mu\nu} - Q^{(\mu\nu)\rho})Q_{\rho\sigma}{}^{\sigma} - Q^{\rho\sigma\mu}Q_{[\rho\sigma]}{}^{\nu}
\end{equation}
and
\begin{equation}
V^{\rho\mu\nu} = \frac{1}{2}Q^{\rho\mu\nu} - Q^{(\mu\nu)\rho} - \frac{1}{2}g^{\mu\nu}(Q^{\rho\sigma}{}_{\sigma} - Q_{\sigma}{}^{\sigma\rho}) + \frac{1}{2}g^{\rho(\mu}Q^{\nu)\sigma}{}_{\sigma} = P^{\rho\mu\nu}\,,
\end{equation}
one finds that they combine into
\begin{multline}
U^{\mu}{}_{\nu} - L^{\sigma}{}_{\sigma\rho}V^{\rho\mu}{}_{\nu} + L^{\rho}{}_{\sigma\rho}V^{\sigma\mu}{}_{\nu} + L^{\mu}{}_{\sigma\rho}V^{\rho\sigma}{}_{\nu} - L^{\sigma}{}_{\nu\rho}V^{\rho\mu}{}_{\sigma}\\
= \frac{1}{2}(Q^{\mu[\rho}{}_{\rho}Q_{\nu\sigma}{}^{\sigma]} - Q_{(\nu}{}^{\mu\rho}Q_{\rho)\sigma}{}^{\rho} + Q^{\rho\mu\sigma}Q_{\nu\rho\sigma}) = -\frac{1}{2}P^{\mu\rho\sigma}Q_{\nu\rho\sigma}\,.
\end{multline}
Combining these results, the metric field equation finally becomes
\begin{equation}\label{eq:fqmetfield}
-\frac{\nabla_{\rho}(\sqrt{-g}f'P^{\rho\mu}{}_{\nu})}{\sqrt{-g}} - \frac{f'}{2}P^{\mu\rho\sigma}Q_{\nu\rho\sigma} + \frac{1}{2}f\delta^{\mu}_{\nu} = \kappa^2\Theta^{\mu}{}_{\nu}\,.
\end{equation}
Note, however, that this form changes if one raises or lowers indices, which appear also under the metric-incompatible covariant derivative \(\nabla_{\rho}\).

Finally, we also take a closer look at the metric teleparallel case, by imposing vanishing nonmetricity. Under this restriction, the general action~\eqref{eq:fgaction} becomes~\cite{Bengochea:2008gz,Linder:2010py,Krssak:2015oua}
\begin{equation}\label{eq:ftaction}
S_{\text{g}} = -\frac{1}{2\kappa^2}\int_Mf(T)\sqrt{-g}\dd^4x\,.
\end{equation}
In this case we need to consider only the single field equation~\eqref{eq:metallfield}, whose left hand side now takes the form
\begin{multline}
W^{\mu\nu} - \nabla_{\rho}Y^{\mu\nu\rho} + Y^{\mu\nu\rho}T^{\tau}{}_{\tau\rho}\\
= \frac{1}{\kappa^2}\left[f'U^{\mu\nu} - \lc{\nabla}_{\rho}(f'V^{\rho\mu\nu}) + \frac{1}{2}fg^{\mu\nu} - \frac{1}{2}\nabla_{\rho}(f'Z^{\mu\nu\rho}) + \frac{1}{2}f'Z^{\mu\nu\rho}T^{\tau}{}_{\tau\rho}\right]\,.
\end{multline}
using the variation expressions~\eqref{eq:fgmetvar} and~\eqref{eq:fgconnvar}. In order to simplify this expressions, we transform the covariant derivative with respect to the independent connection to that of the Levi-Civita connection, and find
\begin{equation}
\nabla_{\rho}(f'Z^{\mu\nu\rho}) - f'Z^{\mu\nu\rho}T^{\tau}{}_{\tau\rho} = \lc{\nabla}_{\rho}(f'Z^{\mu\nu\rho}) + f'(K^{\mu}{}_{\sigma\rho}Z^{\sigma\nu\rho} + K^{\nu}{}_{\sigma\rho}Z^{\mu\sigma\rho})\,,
\end{equation}
where the trace of the torsion tensor cancels with a trace of the contortion tensor. Now we can combine the two covariant derivatives, and evaluate
\begin{equation}
V^{\rho\mu\nu} + \frac{1}{2}Z^{\mu\nu\rho} = 2T_{\sigma}{}^{\sigma[\rho}g^{\nu]\mu} + T^{[\nu\rho]\mu} - \frac{1}{2}T^{\mu\nu\rho} = -S^{\mu\nu\rho}\,.
\end{equation}
We are left with the terms
\begin{equation}
U^{\mu\nu} - \frac{1}{2}(K^{\mu}{}_{\sigma\rho}Z^{\sigma\nu\rho} + K^{\nu}{}_{\sigma\rho}Z^{\mu\sigma\rho}) = 2K^{\nu\rho[\sigma}K_{\rho\sigma}{}^{\mu]} = S^{\rho\sigma\nu}K_{\rho\sigma}{}^{\mu}\,.
\end{equation}
Combining all terms and lowering indices, which commutes with all covariant derivatives since these are now metric-compatible, we can write the field equation as
\begin{equation}\label{eq:ftallfield}
\lc{\nabla}_{\rho}(f'S_{\mu\nu}{}^{\rho}) + f'S^{\rho\sigma}{}_{\nu}K_{\rho\sigma\mu} + \frac{1}{2}fg_{\mu\nu} = \kappa^2(\Theta_{\mu\nu} - \nabla_{\rho}H_{\mu\nu}{}^{\rho} + H_{\mu\nu}{}^{\rho}T^{\tau}{}_{\tau\rho})\,.
\end{equation}
Also this equation can be brought into various other forms by using the identities which hold for the contortion and the torsion.

\subsection{The general quadratic Lagrangians}\label{ssec:quadlag}
The GTEGR action~\eqref{eq:gtegraction} has the appealing property that the gravity scalar~\eqref{eq:gtegrbndterms}, unlike the Ricci scalar, is quadratic in first order derivatives of the dynamical fields, and hence more reminiscent of the kinetic energy of a gauge field. This invites for another class of modified teleparallel gravity theories, by considering an action which is an arbitrary linear combination of all possible scalars which can be obtained by contracting the product of the distortion tensor \(M^{\mu}{}_{\nu\rho}\) with itself. One easily checks that there are 11 possible terms: five terms arise from contracting \(M^{\mu}{}_{\nu\rho}\) with a second copy carrying the same indices in an arbitrary permutation, and six terms arising from contracting two arbitrary traces of the distortion tensor with each other, where in both cases terms which are distinguished only by the order of the factors are counted only once, since they are identical. This gives rise to the generalized gravity scalar~\cite{BeltranJimenez:2019odq}
\begin{equation}\label{eq:gngrscalar}
\begin{split}
\mathcal{G} &= M^{\mu\nu\rho}(k_1M_{\mu\nu\rho} + k_2M_{\nu\rho\mu} + k_3M_{\mu\rho\nu} + k_4M_{\rho\nu\mu} + k_5M_{\nu\mu\rho})\\
&\phantom{=}+ k_6M_{\rho\mu}{}^{\mu}M^{\rho\nu}{}_{\nu} + k_7M_{\mu\rho}{}^{\mu}M^{\nu\rho}{}_{\nu} + k_8M^{\mu}{}_{\mu\rho}M_{\nu}{}^{\nu\rho}\\
&\phantom{=}+ k_9M_{\mu\rho}{}^{\mu}M_{\nu}{}^{\nu\rho} + k_{10}M^{\mu}{}_{\mu\rho}M^{\rho\nu}{}_{\nu} + k_{11}M_{\rho\mu}{}^{\mu}M^{\nu\rho}{}_{\nu}
\end{split}
\end{equation}
with arbitrary constants \(k_1, \ldots, k_{11}\). Equivalently, one could also start from the expression~\eqref{eq:gtegrscalar}, and consider the most general scalar which is quadratic in the torsion and nonmetricity tensors. Again one finds 11 possible terms, so that their most general linear combination is of the form
\begin{equation}\label{eq:gngrtq}
\begin{split}
\mathcal{G} &= a_1T^{\mu\nu\rho}T_{\mu\nu\rho} + a_2T^{\mu\nu\rho}T_{\rho\nu\mu} + a_3T^{\mu}{}_{\mu\rho}T_{\nu}{}^{\nu\rho}\\
&\phantom{=}- b_1Q^{\mu\nu\rho}T_{\rho\nu\mu} - b_2Q^{\rho\mu}{}_{\mu}T^{\nu}{}_{\nu\rho} - b_3Q_{\mu}{}^{\mu\rho}T^{\nu}{}_{\nu\rho}\\
&\phantom{=}+ c_1Q^{\mu\nu\rho}Q_{\mu\nu\rho} + c_2Q^{\mu\nu\rho}Q_{\rho\mu\nu} + c_3Q^{\rho\mu}{}_{\mu}Q_{\rho\nu}{}^{\nu} + c_4Q^{\mu}{}_{\mu\rho}Q_{\nu}{}^{\nu\rho} + c_5Q^{\mu}{}_{\mu\rho}Q^{\rho\nu}{}_{\nu}\,,
\end{split}
\end{equation}
where we introduced the arbitrary constants \(a_1, \ldots, a_3, b_1, \ldots, b_3, c_1, \ldots, c_5\). Demanding that both expressions agree, one easily checks that these two sets of constants are related to each other by
\begin{gather}
k_1 = 2a_1 - b_1 + 2c_1\,, \quad
k_2 = -2a_2 + b_1 + 2c_2\,, \quad
k_9 = -2a_3 + 2b_2 - b_3 + 2c_5\,,\nonumber\\
k_4 = a_2 + c_2\,, \quad
k_5 = a_2 - b_1 + 2c_1\,, \quad
k_6 = c_4\,, \quad
k_7 = a_3 + b_3 + c_4\,,\\
k_8 = a_3 - 2b_2 + 4c_3\,, \quad
k_3 = -2a_1 + b_1 + c_2\,, \quad
k_{10} = -b_3 + 2c_5\,, \quad
k_{11} = b_3 + 2c_4\,.\nonumber
\end{gather}
Further, choosing the values of these constants to be
\begin{equation}\label{eq:gtegrvalues}
k_{11} = -k_2 = 1\,, \quad k_1 = k_3 = k_4 = k_5 = k_6 = k_7 = k_8 = k_9 = k_{10} = 0\,,
\end{equation}
one finds that the scalar \(\mathcal{G}\) reduces to \(G\). Hence, one may expect that the class of modified gravity theories defined by the action
\begin{equation}\label{eq:gngraction}
S_{\text{g}} = -\frac{1}{2\kappa^2}\int_M\mathcal{G}\sqrt{-g}\dd^4x
\end{equation}
has a well-defined limit towards GTEGR, which is achieved if the constant parameters in the action take the aforementioned values. In order to derive the field equations, one can proceed in full analogy to the GTEGR field equations we discussed before. First, it is helpful to calculate the variation of the scalar~\eqref{eq:gngrscalar}, and write it in the form
\begin{equation}\label{eq:gngrvardef}
\delta\mathcal{G} = \mathcal{U}^{\mu\nu}\delta g_{\mu\nu} + \mathcal{V}^{\rho\mu\nu}\lc{\nabla}_{\rho}\delta g_{\mu\nu} + \mathcal{Z}_{\mu}{}^{\nu\rho}\delta\Gamma^{\mu}{}_{\nu\rho}\,.
\end{equation}
Here we have made use of the abbreviations
\begin{subequations}\label{eq:gngrvarabbrev}
\begin{multline}
\mathcal{U}^{\mu\nu} = k_1(M^{\mu\rho\sigma}M^{\nu}{}_{\rho\sigma} - M^{\rho\mu}{}_{\sigma}M_{\rho}{}^{\nu\sigma} - M_{\rho\sigma}{}^{\mu}M^{\rho\sigma\mu}) - k_2M_{\rho}{}^{\sigma(\mu}M_{\sigma}{}^{\nu)\rho}\\
+ k_3(M^{\mu\rho\sigma}M^{\nu}{}_{\sigma\rho} - 2M^{\rho\sigma(\mu}M_{\rho}{}^{\nu)\sigma}) - k_4M^{\rho\mu}{}_{\sigma}M^{\sigma\nu}{}_{\rho} - k_5M^{\rho\sigma\mu}M_{\sigma\rho}{}^{\nu}\\
+ k_6M^{\mu\rho}{}_{\rho}M^{\nu\sigma}{}_{\sigma} - k_7M^{\rho\mu}{}_{\rho}M^{\sigma\nu}{}_{\sigma} - k_8M_{\rho}{}^{\rho\mu}M_{\sigma}{}^{\sigma\nu} - k_9M_{\rho}{}^{\rho(\mu}M_{\sigma}{}^{\nu)\sigma}\\
- (2k_6M_{\rho\sigma}{}^{\sigma} + k_{11}M_{\sigma\rho}{}^{\sigma} + k_{10}M^{\sigma}{}_{\sigma\rho})M^{\rho(\mu\nu)}\,,
\end{multline}
as well as
\begin{multline}
\mathcal{V}^{\rho\mu\nu} = -2k_6g^{\rho(\mu}M^{\nu)\sigma}{}_{\sigma} - k_{11}g^{\rho(\mu}M_{\sigma}{}^{\nu)\sigma} - k_{10}M_{\sigma}{}^{\sigma(\mu}g^{\nu)\rho}\\
+ \frac{1}{2}g^{\mu\nu}\big[(2k_6 - k_{10} - k_{11})M^{\rho\sigma}{}_{\sigma} + (k_{11} - 2k_7 - k_9)M^{\sigma\rho}{}_{\sigma} + (k_{10} - 2k_8 - k_9)M_{\sigma}{}^{\sigma\rho}\big]\\
+ (k_4 - k_5 - k_1 - k_3)M^{(\mu\nu)\rho} + (k_5 - k_4 - k_1 - k_3)M^{(\mu|\rho|\nu)} + (k_1 - k_2 + k_3 - k_4 - k_5)M^{\rho(\mu\nu)}
\end{multline}
and
\begin{multline}
\mathcal{Z}_{\mu}{}^{\nu\rho} = 2k_1M_{\mu}{}^{\nu\rho} + k_2(M^{\nu\rho}{}_{\mu} + M^{\rho}{}_{\mu}{}^{\nu}) + 2k_3M_{\mu}{}^{\rho\nu} + 2k_4M^{\rho\nu}{}_{\mu} + 2k_5M^{\nu}{}_{\mu}{}^{\rho}\\
+ 2k_6M_{\mu\sigma}{}^{\sigma}g^{\nu\rho} + 2k_7M^{\sigma\nu}{}_{\sigma}\delta_{\mu}^{\rho} + 2k_8M_{\sigma}{}^{\sigma\rho}\delta_{\mu}^{\nu} + k_9(M_{\sigma}{}^{\rho\sigma}\delta_{\mu}^{\nu} + M_{\sigma}{}^{\sigma\nu}\delta_{\mu}^{\rho})\\
+ k_{10}(M^{\rho\sigma}{}_{\sigma}\delta_{\mu}^{\nu} + M^{\sigma}{}_{\sigma\mu}g^{\nu\rho}) + k_{11}(M^{\nu\sigma}{}_{\sigma}\delta_{\mu}^{\rho} + M^{\sigma}{}_{\mu\sigma}g^{\nu\rho})\,.
\end{multline}
\end{subequations}
By inserting the variation~\eqref{eq:gngrvardef} into the variation of the action~\eqref{eq:gngraction} and integration by parts, one obtains the form~\eqref{eq:metricgravactvar}, with
\begin{equation}\label{eq:gngrmetvar}
W_{\mu\nu} = \frac{1}{\kappa^2}\left(\mathcal{U}_{\mu\nu} - \lc{\nabla}_{\rho}\mathcal{V}^{\rho}{}_{\mu\nu} + \frac{1}{2}\mathcal{G}g_{\mu\nu}\right)
\end{equation}
and
\begin{equation}\label{eq:gngrconnvar}
Y_{\mu}{}^{\nu\rho} = \frac{1}{2\kappa^2}\mathcal{Z}_{\mu}{}^{\nu\rho}\,.
\end{equation}
Hence, by comparing with the corresponding GTEGR expressions~\eqref{eq:gtegrmetvar} and~\eqref{eq:gtegrconnvar}, we see that these have the same form, and one simply replaces the terms derived by variation of \(G\) with those obtained from \(\mathcal{G}\) in its place. One therefore finds the metric field equation
\begin{equation}\label{eq:gngrmetfield}
\mathcal{U}_{\mu\nu} - \lc{\nabla}_{\rho}(\mathcal{V}^{\rho}{}_{\mu\nu}) + \frac{1}{2}\mathcal{G}g_{\mu\nu} = \kappa^2\Theta_{\mu\nu}\,,
\end{equation}
as well as the connection field equation
\begin{equation}\label{eq:gngrconnfield}
\nabla_{\tau}\mathcal{Z}_{\mu}{}^{\nu\tau} - M^{\omega}{}_{\tau\omega}\mathcal{Z}_{\mu}{}^{\nu\tau} = 2\kappa^2(\nabla_{\tau}H_{\mu}{}^{\nu\tau} - M^{\omega}{}_{\tau\omega}H_{\mu}{}^{\nu\tau})\,,
\end{equation}
with the abbreviations~\eqref{eq:gngrvarabbrev}.

A more comprehensible set of field equations is obtained for the more restricted geometries, in which we impose either vanishing torsion or vanishing nonmetricity. This can most easily be seen from the expression~\eqref{eq:gngrtq}, which shows that numerous terms vanish identically in either of these two cases. We first consider the symmetric teleparallel case of vanishing torsion. In this case, \(\mathcal{G}\) reduces to the generalized nonmetricity scalar~\cite{BeltranJimenez:2017tkd}
\begin{equation}\label{eq:sngrscalar}
\begin{split}
\mathcal{Q} &= \frac{1}{2}Q_{\rho\mu\nu}\mathcal{P}^{\rho\mu\nu}\\
&= c_1Q^{\mu\nu\rho}Q_{\mu\nu\rho} + c_2Q^{\mu\nu\rho}Q_{\rho\mu\nu} + c_3Q^{\rho\mu}{}_{\mu}Q_{\rho\nu}{}^{\nu} + c_4Q^{\mu}{}_{\mu\rho}Q_{\nu}{}^{\nu\rho} + c_5Q^{\mu}{}_{\mu\rho}Q^{\rho\nu}{}_{\nu}\,,
\end{split}
\end{equation}
and only the five constant parameters \(c_1, \ldots, c_5\) remain present in the action. In place of the nonmetricity conjugate~\eqref{eq:nonmetconj} we now have the generalized expression
\begin{multline}\label{eq:ngrnonmetconj}
\mathcal{P}^{\rho\mu\nu} = 2c_1Q^{\rho\mu\nu} + 2c_2Q^{(\mu\nu)\rho} + 2c_3g^{\mu\nu}Q^{\rho\sigma}{}_{\sigma}\\
+ 2c_4Q_{\sigma}{}^{\sigma(\mu}g^{\nu)\rho} + c_5(g^{\mu\nu}Q_{\sigma}{}^{\sigma\rho} + g^{\rho(\mu}Q^{\nu)\sigma}{}_{\sigma})\,.
\end{multline}
For the corresponding class of gravity theories depending on these parameters, whose action reads
\begin{equation}\label{eq:sngraction}
S_{\text{g}} = -\frac{1}{2\kappa^2}\int_M\mathcal{Q}\sqrt{-g}\dd^4x\,,
\end{equation}
the term ``Newer General Relativity'' has been coined. Its field equations can be obtained in great analogy to the other symmetric teleparallel gravity theories we have encountered before. First, we derive the connection field equation~\eqref{eq:symconnfield}, and use the fact that only the symmetric part \(Y_{\mu}{}^{(\nu\rho)}\) contributes. Using the variation~\eqref{eq:gngrconnvar}, we thus calculate
\begin{multline}
\mathcal{Z}_{\mu}{}^{(\nu\rho)} = -2c_2Q_{\mu}{}^{\nu\rho} - 2(2c_1 + c_2)Q^{(\nu\rho)}{}_{\mu} - 2g^{\nu\rho}(2c_4Q^{\sigma}{}_{\sigma\mu} + c_5Q_{\mu\sigma}{}^{\sigma})\\
- 4(c_4 + c_5)Q_{\sigma}{}^{\sigma(\nu}\delta^{\rho)}_{\mu} - 2(4c_3 + c_5)\delta^{(\nu}_{\mu}Q^{\rho)\sigma}{}_{\sigma} = -2\mathcal{P}^{(\nu\rho)}{}_{\mu}\,,
\end{multline}
which generalizes the similar relation~\eqref{eq:symconnvarrel}. Hence, we find that the connection field equation can be written in the simple form
\begin{equation}
-\nabla_{\nu}\nabla_{\rho}(\tilde{\mathcal{P}}^{\nu\rho}{}_{\mu}) = \kappa^2\nabla_{\nu}\nabla_{\rho}\tilde{H}_{\mu}{}^{\nu\rho}\,,
\end{equation}
using the tensor density \(\tilde{\mathcal{P}}^{\nu\rho}{}_{\mu}\) built from the generalized nonmetricity conjugate~\eqref{eq:ngrnonmetconj}. We then proceed with the metric equation, which still takes the general form~\eqref{eq:gngrmetfield} also in the symmetric teleparallel case, but can be simplified as follows. Raising one index and introducing a density factor, it can equivalently be written as
\begin{equation}
\mathcal{U}^{\mu}{}_{\nu} - \frac{\lc{\nabla}_{\rho}(\sqrt{-g}\mathcal{V}^{\rho\mu}{}_{\nu})}{\sqrt{-g}} + \frac{1}{2}\mathcal{G}\delta^{\mu}_{\nu} = \kappa^2\Theta^{\mu}{}_{\nu}\,.
\end{equation}
The covariant derivative with respect to the Levi-Civita connection can be transformed to the independent connection, by using the relation
\begin{multline}
\lc{\nabla}_{\rho}(\sqrt{-g}\mathcal{V}^{\rho\mu}{}_{\nu}) = \nabla_{\rho}(\sqrt{-g}\mathcal{V}^{\rho\mu}{}_{\nu})\\
+ \sqrt{-g}(L^{\sigma}{}_{\sigma\rho}\mathcal{V}^{\rho\mu}{}_{\nu} - L^{\rho}{}_{\sigma\rho}\mathcal{V}^{\sigma\mu}{}_{\nu} - L^{\mu}{}_{\sigma\rho}\mathcal{V}^{\rho\sigma}{}_{\nu} + L^{\sigma}{}_{\nu\rho}\mathcal{V}^{\rho\mu}{}_{\sigma})\,.
\end{multline}
To proceed further, we need the terms
\begin{multline}
\mathcal{U}^{\mu\nu} = (2c_1Q^{\rho\sigma\mu} + c_2Q^{\sigma\rho\mu})Q_{\rho\sigma}{}^{\nu} - (2c_1 + c_2)Q^{\rho\sigma(\mu}Q^{\nu)}{}_{\rho\sigma} - (c_1 + c_2)Q^{\mu\rho\sigma}Q^{\nu}{}_{\rho\sigma}\\
- c_4Q_{\rho}{}^{\rho(\mu}Q^{\nu)\sigma}{}_{\sigma} + 2c_4Q_{\rho}{}^{\rho\mu}Q_{\sigma}{}^{\sigma\nu} - \left(c_3 + \frac{c_5}{2}\right)Q^{\mu\rho}{}_{\rho}Q^{\nu\sigma}{}_{\sigma}\\
+ (2c_4Q^{\sigma}{}_{\sigma\rho} + c_5Q_{\rho\sigma}{}^{\sigma})\left(\frac{1}{2}Q^{\rho\mu\nu} - Q^{(\mu\nu)\rho}\right)
\end{multline}
and
\begin{multline}
\mathcal{V}^{\rho\mu\nu} = 2c_1Q^{\rho\mu\nu} + 2c_2Q^{(\mu\nu)\rho} + 2c_3g^{\mu\nu}Q^{\rho\sigma}{}_{\sigma}\\
+ 2c_4Q_{\sigma}{}^{\sigma(\mu}g^{\nu)\rho} + c_5(g^{\mu\nu}Q_{\sigma}{}^{\sigma\rho} + g^{\rho(\mu}Q^{\nu)\sigma}{}_{\sigma})\,,
\end{multline}
which are obtained from the more general expressions~\eqref{eq:gngrvarabbrev} by imposing vanishing torsion. A tedious, but straightforward calculation shows that the resulting terms can be combined to yield
\begin{multline}
\mathcal{U}^{\mu}{}_{\nu} - L^{\sigma}{}_{\sigma\rho}\mathcal{V}^{\rho\mu}{}_{\nu} + L^{\rho}{}_{\sigma\rho}\mathcal{V}^{\sigma\mu}{}_{\nu} + L^{\mu}{}_{\sigma\rho}\mathcal{V}^{\rho\sigma}{}_{\nu} - L^{\sigma}{}_{\nu\rho}\mathcal{V}^{\rho\mu}{}_{\sigma}\\
= -(c_1Q^{\mu\rho\sigma} + c_2Q^{\rho\sigma\mu})Q_{\nu\rho\sigma} - c_3Q^{\mu\rho}{}_{\rho}Q_{\nu\sigma}{}^{\sigma} - c_4Q^{\rho}{}_{\rho\sigma}Q_{\nu}{}^{\mu\sigma} - c_5Q_{(\nu}{}^{\mu\rho}Q_{\rho)\sigma}{}^{\sigma}\\
= -\frac{1}{2}\mathcal{P}^{\mu\rho\sigma}Q_{\nu\rho\sigma}\,.
\end{multline}
This finally yields the metric field equation
\begin{equation}
-\frac{\nabla_{\rho}(\sqrt{-g}\mathcal{P}^{\rho\mu}{}_{\nu})}{\sqrt{-g}} - \frac{1}{2}\mathcal{P}^{\mu\rho\sigma}Q_{\nu\rho\sigma} + \frac{1}{2}\mathcal{Q}\delta^{\mu}_{\nu} = \kappa^2\Theta^{\mu}{}_{\nu}
\end{equation}
for the Newer General Relativity class of gravity theories, where we now also used the relation \(\mathcal{G} = \mathcal{Q}\) in the absence of torsion. Note that a special case is obtained when the parameters take the values~\eqref{eq:gtegrvalues}, for which we have
\begin{equation}
c_1 = \frac{1}{4}\,, \quad
c_2 = -\frac{1}{2}\,, \quad
c_3 = -\frac{1}{4}\,, \quad
c_4 = 0\,, \quad
c_5 = \frac{1}{2}\,.
\end{equation}
In this case we find \(\mathcal{Q} = Q\) and \(\mathcal{P}^{\mu\nu\rho} = P^{\mu\nu\rho}\), so that the theory reduces to STEGR.

Finally, also in the metric teleparallel geometry we can find a general class of gravity theories, whose action is now quadratic in the torsion tensor. By imposing vanishing nonmetricity, the scalar~\eqref{eq:gngrscalar} becomes the generalized torsion scalar~\cite{Hayashi:1979qx}
\begin{equation}\label{eq:mngrscalar}
\begin{split}
\mathcal{T} &= \frac{1}{2}T^{\rho}{}_{\mu\nu}\mathcal{S}_{\rho}{}^{\mu\nu}\\
&= a_1T^{\mu\nu\rho}T_{\mu\nu\rho} + a_2T^{\mu\nu\rho}T_{\rho\nu\mu} + a_3T^{\mu}{}_{\mu\rho}T_{\nu}{}^{\nu\rho}\,,
\end{split}
\end{equation}
where the generalized superpotential is now given by
\begin{equation}\label{eq:ngrsuppot}
\mathcal{S}_{\rho}{}^{\mu\nu} = 2a_1T_{\rho}{}^{\mu\nu} + 2a_2T^{[\nu\mu]}{}_{\rho} + 2a_3T_{\sigma}{}^{\sigma[\nu}\delta^{\mu]}_{\rho}\,.
\end{equation}
The resulting class of gravity theories, which is now defined by the action
\begin{equation}\label{eq:mngraction}
S_{\text{g}} = -\frac{1}{2\kappa^2}\int_M\mathcal{T}\sqrt{-g}\dd^4x\,,
\end{equation}
is known as ``New General Relativity''\footnote{This term is also, more commonly, used for a particular subclass of theories, in which \(2a_1 + a_2 = 0\) and \(a_3 = -1\), so that there is only one free parameter besides the gravitational constant \(\kappa\)~\cite{Hayashi:1979qx}.}. In this case, the left hand side of the field equations~\eqref{eq:metallfield} becomes
\begin{multline}
W^{\mu\nu} - \nabla_{\rho}Y^{\mu\nu\rho} + Y^{\mu\nu\rho}T^{\tau}{}_{\tau\rho}\\
= \frac{1}{\kappa^2}\left[\mathcal{U}^{\mu\nu} - \lc{\nabla}_{\rho}\mathcal{V}^{\rho\mu\nu} + \frac{1}{2}\mathcal{G}g^{\mu\nu} - \frac{1}{2}\nabla_{\rho}\mathcal{Z}^{\mu\nu\rho} + \frac{1}{2}\mathcal{Z}^{\mu\nu\rho}T^{\tau}{}_{\tau\rho}\right]\,.
\end{multline}
with the help of the formulas~\eqref{eq:gngrmetvar} and~\eqref{eq:gngrconnvar}. In order to combine the two derivative terms, we convert the covariant derivative \(\nabla_{\rho}\) with respect to the independent connection to a Levi-Civita covariant derivative \(\lc{\nabla}_{\rho}\), using the relation
\begin{equation}
\nabla_{\rho}\mathcal{Z}^{\mu\nu\rho} - \mathcal{Z}^{\mu\nu\rho}T^{\tau}{}_{\tau\rho} = \lc{\nabla}_{\rho}\mathcal{Z}^{\mu\nu\rho} + K^{\mu}{}_{\sigma\rho}\mathcal{Z}^{\sigma\nu\rho} + K^{\nu}{}_{\sigma\rho}\mathcal{Z}^{\mu\sigma\rho}\,.
\end{equation}
Now the two terms under the derivative combine into
\begin{equation}
\mathcal{V}^{\rho\mu\nu} + \frac{1}{2}\mathcal{Z}^{\mu\nu\rho} = -2a_1T^{\mu\nu\rho} - 2a_2T^{[\rho\nu]\mu} - 2a_3T_{\sigma}{}^{\sigma[\rho}g^{\nu]\mu} = -\mathcal{S}^{\mu\nu\rho}\,.
\end{equation}
The remaining terms take, once again, a very simple form, which is given by
\begin{multline}
\mathcal{U}^{\mu\nu} - \frac{1}{2}(K^{\mu}{}_{\sigma\rho}\mathcal{Z}^{\sigma\nu\rho} + K^{\nu}{}_{\sigma\rho}\mathcal{Z}^{\mu\sigma\rho})\\
= [(a_2 - 2a_1)K^{\rho\sigma\nu} + (3a_2 - 2a_1)K^{\nu\rho\sigma}]K_{\rho\sigma}{}^{\mu} + a_3K_{\rho\sigma}{}^{\sigma}K^{\nu\rho\mu} = \mathcal{S}^{\rho\sigma\nu}K_{\rho\sigma}{}^{\mu}\,.
\end{multline}
Hence, the full field equations of New General Relativity become
\begin{equation}
\lc{\nabla}_{\rho}(\mathcal{S}_{\mu\nu}{}^{\rho}) + \mathcal{S}^{\rho\sigma}{}_{\nu}K_{\rho\sigma\mu} + \frac{1}{2}\mathcal{G}g_{\mu\nu} = \kappa^2(\Theta_{\mu\nu} - \nabla_{\rho}H_{\mu\nu}{}^{\rho} + H_{\mu\nu}{}^{\rho}T^{\tau}{}_{\tau\rho})\,.
\end{equation}
Also for this class of theories a special case is obtained by choosing the parameter values~\eqref{eq:gtegrvalues}, which now implies
\begin{equation}
a_1 = \frac{1}{4}\,, \quad
a_2 = \frac{1}{2}\,, \quad
a_3 = -1\,.
\end{equation}
In this case, the theory reduces to MTEGR, with \(\mathcal{T} = T\) and \(\mathcal{S}_{\rho}{}^{\mu\nu} = S_{\rho}{}^{\mu\nu}\).

\subsection{Scalar-teleparallel theories}\label{ssec:scalartele}
While the classes of modified teleparallel gravity theories we considered so far were constructed purely from the metric and the flat affine connection as fundamental fields, we now consider a class of theories in which in addition a scalar field is introduced as a fundamental field variable. Also this class of theories can be motivated by analogy with a scalar-tensor modification of the Einstein-Hilbert action~\eqref{eq:einsteinhilbert} of general relativity, which takes the general form
\begin{equation}\label{eq:stcaction}
S_{\text{g}} = \frac{1}{2\kappa^2}\int_M\left[\mathcal{A}(\phi)\lc{R} - \mathcal{B}(\phi)g^{\mu\nu}\lc{\nabla}_{\mu}\phi\lc{\nabla}_{\nu}\phi - 2\kappa^2\mathcal{V}(\phi)\right]\sqrt{-g}\dd^4x\,,
\end{equation}
where \(\mathcal{A}, \mathcal{B}, \mathcal{V}\) are free functions of the scalar field \(\phi\). Here we work in the so-called Jordan frame, which means that we assume no direct coupling between the scalar field and any matter fields. Recalling that the Ricci scalar \(\lc{R}\) can be written in the form~\eqref{eq:riccisplit}, one may expect that replacing \(\lc{R}\) by \(-G + B\) one obtains a teleparallel equivalent of the scalar-curvature theory, while using only \(-G\) instead leads to an inequivalent scalar-teleparallel theory, since the omitted term is not a boundary term due to the non-minimal coupling term \(\mathcal{A}(\phi)\). One can cover both cases by considering the action
\begin{equation}
S_{\text{g}} = \frac{1}{2\kappa^2}\int_M\left[-\mathcal{A}(\phi)G - \mathcal{B}(\phi)g^{\mu\nu}\lc{\nabla}_{\mu}\phi\lc{\nabla}_{\nu}\phi - \hat{\mathcal{C}}(\phi)B - 2\kappa^2\mathcal{V}(\phi)\right]\sqrt{-g}\dd^4x\,,
\end{equation}
where we introduced another free function \(\hat{\mathcal{C}}\) of the scalar field. Keeping in mind that \(B\) is a boundary term, i.e., a total divergence, we see that the field equations do not change if we add an arbitrary constant to \(\hat{\mathcal{C}}\). To resolve this ambiguity, we can use integration by parts,
\begin{equation}
2\hat{\mathcal{C}}\lc{\nabla}_{\mu}M^{[\nu\mu]}{}_{\nu} = \lc{\nabla}_{\mu}(\hat{\mathcal{C}}M^{[\nu\mu]}{}_{\nu}) - \hat{\mathcal{C}}'M^{[\nu\mu]}{}_{\nu}\lc{\nabla}_{\mu}\phi\,,
\end{equation}
and omit the boundary term. Defining a new parameter function \(\mathcal{C} = \hat{\mathcal{C}}'\), we then have
\begin{multline}\label{eq:stgaction}
S_{\text{g}} = \frac{1}{2\kappa^2}\int_M\Big[-\mathcal{A}(\phi)G - \mathcal{B}(\phi)g^{\mu\nu}\lc{\nabla}_{\mu}\phi\lc{\nabla}_{\nu}\phi\\
+ 2\mathcal{C}(\phi)M^{[\nu\mu]}{}_{\nu}\lc{\nabla}_{\mu}\phi - 2\kappa^2\mathcal{V}(\phi)\Big]\sqrt{-g}\dd^4x\,.
\end{multline}
Note that for \(\mathcal{A}' + \mathcal{C} = 0\), the action becomes equivalent to the scalar-curvature action~\eqref{eq:stcaction}. To derive the field equations for this generalized class of theories, we proceed by varying the action as with the previous examples. Due to the presence of an additional fundamental field, also the variation~\eqref{eq:metricgravactvar} is enhanced by an additional term, and becomes
\begin{equation}
\delta S_{\text{g}} = -\int_M\left(\frac{1}{2}W^{\mu\nu}\delta g_{\mu\nu} + Y_{\mu}{}^{\nu\rho}\delta\Gamma^{\mu}{}_{\nu\rho} + \Phi\delta\phi\right)\sqrt{-g}\dd^4x\,,
\end{equation}
after eliminating derivatives of the variations using integration by parts. Varying the action~\eqref{eq:stgaction}, we find the terms
\begin{subequations}\label{eq:stgvar}
\begin{align}
W_{\mu\nu} &= \frac{1}{\kappa^2}\bigg\{\mathcal{A}\lc{R}_{\mu\nu} - \frac{\mathcal{A}}{2}\lc{R}g_{\mu\nu} + \mathcal{C}\lc{\nabla}_{\mu}\lc{\nabla}_{\nu}\phi - (\mathcal{B} - \mathcal{C}')\lc{\nabla}_{\mu}\phi\lc{\nabla}_{\nu}\phi\nonumber\\
&\phantom{=}+ (\mathcal{A}' + \mathcal{C})\left(\lc{\nabla}_{(\mu}\phi M^{\rho}{}_{\nu)\rho} - M^{\rho}{}_{(\mu\nu)}\lc{\nabla}_{\rho}\phi + M^{[\rho\sigma]}{}_{\sigma}\lc{\nabla}_{\rho}\phi g_{\mu\nu}\right)\nonumber\\
&\phantom{=}+ \left[\left(\frac{\mathcal{B}}{2} - \mathcal{C}'\right)\lc{\nabla}_{\rho}\phi\lc{\nabla}^{\rho}\phi - \mathcal{C}\lc{\nabla}_{\rho}\lc{\nabla}^{\rho}\phi + \kappa^2\mathcal{V}\right]g_{\mu\nu}\bigg\}\,,\\
Y_{\mu}{}^{\nu\rho} &= \frac{1}{2\kappa^2}\left[\mathcal{A}(g^{\nu\rho}M^{\sigma}{}_{\mu\sigma} + \delta_{\mu}^{\rho}M^{\nu\sigma}{}_{\sigma} - M^{\nu\rho}{}_{\mu} - M^{\rho}{}_{\nu}{}^{\mu}) + \mathcal{C}(g^{\nu\rho}\lc{\nabla}_{\mu}\phi - \delta_{\mu}^{\rho}\lc{\nabla}^{\nu}\phi)\right]\,,\\
\Phi &= \frac{1}{2\kappa^2}\left[-2\mathcal{B}\lc{\nabla}_{\mu}\lc{\nabla}^{\mu}\phi - \mathcal{B}'\lc{\nabla}_{\mu}\phi\lc{\nabla}^{\mu}\phi + \mathcal{C}B + \mathcal{A}'G\right] + \mathcal{V}\,,
\end{align}
\end{subequations}
where we have made use of the relations~\eqref{eq:curvdec} and~\eqref{eq:gtegrbndterms}, and from now on we omit the argument \(\phi\) of the parameter functions for brevity. We can then read off the field equations and study their properties. We start with the metric field equation~\eqref{eq:genmetfield}, which reads
\begin{multline}\label{eq:stgmetfield}
\mathcal{A}\lc{R}_{\mu\nu} - \frac{\mathcal{A}}{2}\lc{R}g_{\mu\nu} + \mathcal{C}\lc{\nabla}_{\mu}\lc{\nabla}_{\nu}\phi + (\mathcal{A}' + \mathcal{C})\left(\lc{\nabla}_{(\mu}\phi M^{\rho}{}_{\nu)\rho} - M^{\rho}{}_{(\mu\nu)}\lc{\nabla}_{\rho}\phi + M^{[\rho\sigma]}{}_{\sigma}\lc{\nabla}_{\rho}\phi g_{\mu\nu}\right)\\
- (\mathcal{B} - \mathcal{C}')\lc{\nabla}_{\mu}\phi\lc{\nabla}_{\nu}\phi + \left[\left(\frac{\mathcal{B}}{2} - \mathcal{C}'\right)\lc{\nabla}_{\rho}\phi\lc{\nabla}^{\rho}\phi - \mathcal{C}\lc{\nabla}_{\rho}\lc{\nabla}^{\rho}\phi + \kappa^2\mathcal{V}\right]g_{\mu\nu} = \kappa^2\Theta_{\mu\nu}\,.
\end{multline}
It is most remarkable that in the case \(\mathcal{A}' + \mathcal{C} = 0\) the only term containing the flat, affine connection vanishes from these field equations, and one finds that they indeed resemble the field equations of scalar-curvature gravity in this case. To check whether this property holds also for the connection field equation~\eqref{eq:genconnfield}, we calculate
\begin{equation}
\nabla_{\tau}Y_{\mu}{}^{\nu\tau} - M^{\omega}{}_{\tau\omega}Y_{\mu}{}^{\nu\tau} = \frac{\mathcal{A}' + \mathcal{C}}{2\kappa^2}\left[M^{\nu\rho}{}_{\rho}\lc{\nabla}_{\mu}\phi + M^{\rho}{}_{\mu\rho}\lc{\nabla}^{\nu}\phi - (M^{\nu\rho}{}_{\mu} + M^{\rho}{}_{\mu}{}^{\nu})\lc{\nabla}_{\rho}\phi\right]\,,
\end{equation}
where any terms involving the covariant derivative of the distortion \(M^{\mu}{}_{\nu\rho}\) cancel as a consequence of the flatness of the connection. We see that this expression becomes trivial for \(\mathcal{A}' + \mathcal{C} = 0\). In that case, the connection field equation
\begin{multline}\label{eq:stgconnfield}
(\mathcal{A}' + \mathcal{C})\left[M^{\nu\rho}{}_{\rho}\lc{\nabla}_{\mu}\phi + M^{\rho}{}_{\mu\rho}\lc{\nabla}^{\nu}\phi - (M^{\nu\rho}{}_{\mu} + M^{\rho}{}_{\mu}{}^{\nu})\lc{\nabla}_{\rho}\phi\right]\\
= 2\kappa^2(\nabla_{\tau}H_{\mu}{}^{\nu\tau} - M^{\omega}{}_{\tau\omega}H_{\mu}{}^{\nu\tau})
\end{multline}
becomes a constraint for the hypermomentum. Finally, we study the scalar field equation
\begin{equation}
-2\mathcal{B}\lc{\nabla}_{\mu}\lc{\nabla}^{\mu}\phi - \mathcal{B}'\lc{\nabla}_{\mu}\phi\lc{\nabla}^{\mu}\phi + \mathcal{C}B + \mathcal{A}'G + 2\kappa^2\mathcal{V}' = 0\,.
\end{equation}
Here the right hand side vanishes, since we do not consider any direct coupling between the scalar field and matter. Note that if \(\mathcal{C} = -\mathcal{A}'\), i.e., in the case of the scalar-curvature equivalent, the two terms \(\mathcal{C}B + \mathcal{A}'G\) combine to \(-\mathcal{A}'\lc{R}\), and the equation becomes independent of the teleparallel connection, as one would expect, and as we have seen for the remaining field equations. Further, one finds that the scalar field equation contains second order derivatives of both the scalar field and the metric, where the latter enter through the boundary term. In order to eliminate these metric derivatives from the equation, it is common to apply a ``debraiding'' procedure by adding a suitable multiple of the trace of the matter field equation. The latter reads
\begin{multline}
-\mathcal{A}\lc{R} - 3\mathcal{C}\lc{\nabla}_{\mu}\lc{\nabla}^{\mu}\phi + 2(\mathcal{A}' + \mathcal{C})M^{[\mu\nu]}{}_{\nu}\lc{\nabla}_{\mu}\phi\\
+ (\mathcal{B} - 3\mathcal{C}')\lc{\nabla}_{\mu}\phi\lc{\nabla}^{\mu}\phi + 4\kappa^2\mathcal{V} = \kappa^2\Theta_{\mu}{}^{\mu}\,.
\end{multline}
Hence, calculating the linear combination
\begin{multline}
\mathcal{C}W_{\mu}{}^{\mu} + 2\mathcal{A}\Phi = \frac{1}{\kappa^2}\bigg[-(2\mathcal{A}\mathcal{B} + 3\mathcal{C}^2)\lc{\nabla}_{\mu}\lc{\nabla}^{\mu}\phi + (\mathcal{B}\mathcal{C} - 3\mathcal{C}\mathcal{C}' - \mathcal{A}\mathcal{B}')\lc{\nabla}_{\mu}\phi\lc{\nabla}^{\mu}\phi\\
+ (\mathcal{A}' + \mathcal{C})\left(\mathcal{A}G + 2\mathcal{C}M^{[\mu\nu]}{}_{\nu}\lc{\nabla}_{\mu}\phi\right)\bigg] + 2\mathcal{A}\mathcal{V}' + 4\mathcal{C}\mathcal{V}\,,
\end{multline}
we find that the debraided scalar field equation
\begin{multline}\label{eq:genscaldebfield}
-(2\mathcal{A}\mathcal{B} + 3\mathcal{C}^2)\lc{\nabla}_{\mu}\lc{\nabla}^{\mu}\phi + (\mathcal{B}\mathcal{C} - 3\mathcal{C}\mathcal{C}' - \mathcal{A}\mathcal{B}')\lc{\nabla}_{\mu}\phi\lc{\nabla}^{\mu}\phi\\
+ (\mathcal{A}' + \mathcal{C})\left(\mathcal{A}G + 2\mathcal{C}M^{[\mu\nu]}{}_{\nu}\lc{\nabla}_{\mu}\phi\right) + 2\kappa^2(\mathcal{A}\mathcal{V}' + 2\mathcal{C}\mathcal{V}) = \kappa^2\mathcal{C}\Theta_{\mu}{}^{\mu}
\end{multline}
does not contain any derivatives of the independent connection, and has only first order derivatives of the metric tensor, which enter through the distortion and the Christoffel symbols contained in the covariant derivative. Also here we see that the teleparallel connection does not contribute to the field equation for \(\mathcal{A}' + \mathcal{C} = 0\). Further, one finds that the trace of the energy-momentum tensor acts as the matter source for the scalar field.

It is now easy to study how the field equations change if we consider the symmetric or metric teleparallel geometries instead of the general teleparallel geometry we have used to construct the scalar-teleparallel gravity theory discussed above. We start with the former, which yields a class of scalar-nonmetricity theories of gravity, whose action is given by~\cite{Jarv:2018bgs,Runkla:2018xrv}
\begin{multline}\label{eq:stqaction}
S_{\text{g}} = \frac{1}{2\kappa^2}\int_M\Big[-\mathcal{A}(\phi)Q - \mathcal{B}(\phi)g^{\mu\nu}\lc{\nabla}_{\mu}\phi\lc{\nabla}_{\nu}\phi\\
+ \mathcal{C}(\phi)(Q_{\nu}{}^{\nu\mu} - Q^{\mu\nu}{}_{\nu})\lc{\nabla}_{\mu}\phi - 2\kappa^2\mathcal{V}(\phi)\Big]\sqrt{-g}\dd^4x\,.
\end{multline}
For the metric field equations, which retain the general form~\eqref{eq:genmetfield}, we see that the only change compared to the general teleparallel case arises from those terms which involve the teleparallel affine connection. These terms greatly simplify and become
\begin{equation}
\lc{\nabla}_{(\mu}\phi M^{\rho}{}_{\nu)\rho} - M^{\rho}{}_{(\mu\nu)}\lc{\nabla}_{\rho}\phi + M^{[\rho\sigma]}{}_{\sigma}\lc{\nabla}_{\rho}\phi g_{\mu\nu} = -P^{\rho}{}_{\mu\nu}\lc{\nabla}_{\rho}\phi\,,
\end{equation}
using the nonmetricity conjugate~\eqref{eq:nonmetconj}. The metric field equations therefore read
\begin{multline}\label{eq:stqmetfield}
\mathcal{A}\lc{R}_{\mu\nu} - \frac{\mathcal{A}}{2}\lc{R}g_{\mu\nu} + \mathcal{C}\lc{\nabla}_{\mu}\lc{\nabla}_{\nu}\phi - (\mathcal{B} - \mathcal{C}')\lc{\nabla}_{\mu}\phi\lc{\nabla}_{\nu}\phi - (\mathcal{A}' + \mathcal{C})P^{\rho}{}_{\mu\nu}\lc{\nabla}_{\rho}\phi\\
+ \left[\left(\frac{\mathcal{B}}{2} - \mathcal{C}'\right)\lc{\nabla}_{\rho}\phi\lc{\nabla}^{\rho}\phi - \mathcal{C}\lc{\nabla}_{\rho}\lc{\nabla}^{\rho}\phi + \kappa^2\mathcal{V}\right]g_{\mu\nu} = \kappa^2\Theta_{\mu\nu}\,.
\end{multline}
We then continue with the connection equation, which now takes the form~\eqref{eq:symconnfield}. Here we can make use of several simplifications we have employed before. First, using the variation~\eqref{eq:gvarabbrev} of the gravity scalar \(G\), we write the variation~\eqref{eq:stgvar} as
\begin{equation}
\begin{split}
Y_{\mu}{}^{\nu\rho} &= \frac{1}{2\kappa^2}\left[\mathcal{A}Z_{\mu}{}^{\nu\rho} + \mathcal{C}\left(g^{\nu\rho}\lc{\nabla}_{\mu}\phi - \delta_{\mu}^{\rho}\lc{\nabla}^{\nu}\phi\right)\right]\\
&= \frac{1}{2\kappa^2}\left[\mathcal{A}Z_{\mu}{}^{\nu\rho} + \mathcal{C}\left(g^{\nu\rho}\nabla_{\mu}\phi - \delta_{\mu}^{\rho}g^{\nu\sigma}\nabla_{\sigma}\phi\right)\right]\,,
\end{split}
\end{equation}
where we used the fact that any covariant derivative acts equally on the scalar field \(\phi\). Next, we introduce a density factor \(\sqrt{-g}\) and take a covariant derivative, to calculate
\begin{equation}
\begin{split}
\nabla_{\rho}\tilde{Y}_{\mu}{}^{\nu\rho} &= \frac{1}{2\kappa^2}\nabla_{\rho}\left[\mathcal{A}\tilde{Z}_{\mu}{}^{\nu\rho} + \sqrt{-g}\mathcal{C}\left(g^{\nu\rho}\nabla_{\mu}\phi - \delta_{\mu}^{\rho}g^{\nu\sigma}\nabla_{\sigma}\phi\right)\right]\\
&= \frac{\sqrt{-g}}{2\kappa^2}\bigg[\mathcal{A}'Z_{\mu}{}^{\nu\rho}\nabla_{\rho}\phi + \left(\frac{1}{2}Q_{\rho\tau}{}^{\tau}\mathcal{C} + \mathcal{C}'\nabla_{\rho}\phi\right)\left(g^{\nu\rho}\nabla_{\mu}\phi - \delta_{\mu}^{\rho}g^{\nu\sigma}\nabla_{\sigma}\phi\right)\\
&\phantom{=}+ \mathcal{C}\left(g^{\nu\rho}\nabla_{\rho}\nabla_{\mu}\phi - g^{\nu\sigma}\nabla_{\mu}\nabla_{\sigma}\phi - Q_{\rho}{}^{\rho\nu}\nabla_{\mu}\phi + Q_{\mu}{}^{\nu\sigma}\nabla_{\sigma}\phi\right)\bigg]\\
&= \frac{1}{2\kappa^2}(\mathcal{A}' + \mathcal{C})\tilde{Z}_{\mu}{}^{\nu\rho}\nabla_{\rho}\phi\\
&= \frac{1}{2\kappa^2}\nabla_{\rho}[(\mathcal{A} + \hat{\mathcal{C}})\tilde{Z}_{\mu}{}^{\nu\rho}]\,,
\end{split}
\end{equation}
where we used the identity \(\nabla_{\rho}\tilde{Z}_{\mu}{}^{\nu\rho} = 0\) we found in deriving the STEGR field equations, and the fact that numerous terms involving the scalar field cancel, while the remaining terms combine to a very compact form. Here \(\hat{\mathcal{C}}\) is defined by \(\hat{\mathcal{C}}' = \mathcal{C}\) only up to an irrelevant constant. To obtain the connection field equations, we apply another covariant derivative, and use the relation~\eqref{eq:symconnvarrel} to finally obtain
\begin{equation}
\nabla_{\nu}\nabla_{\rho}\tilde{Y}_{\mu}{}^{\nu\rho} = \frac{1}{2\kappa^2}\nabla_{\nu}\nabla_{\rho}[(\mathcal{A} + \hat{\mathcal{C}})\tilde{Z}_{\mu}{}^{\nu\rho}] = -\frac{1}{\kappa^2}\nabla_{\nu}\nabla_{\rho}[(\mathcal{A} + \hat{\mathcal{C}})\tilde{P}^{\nu\rho}{}_{\mu}]\,.
\end{equation}
Hence, we see that the left hand side of the connection field equations
\begin{equation}\label{eq:stqconnfield}
-\nabla_{\nu}\nabla_{\rho}[(\mathcal{A} + \hat{\mathcal{C}})\tilde{P}^{\nu\rho}{}_{\mu}] = \kappa^2\nabla_{\nu}\nabla_{\rho}\tilde{H}_{\mu}{}^{\nu\rho}
\end{equation}
vanishes identically for \(\mathcal{A}' + \mathcal{C} = 0\). At last, we come to the scalar field equation, which we consider in its debraided form~\eqref{eq:genscaldebfield}. Imposing vanishing torsion, the only affected term is given by
\begin{equation}
\mathcal{A}G + 2\mathcal{C}M^{[\mu\nu]}{}_{\nu}\lc{\nabla}_{\mu}\phi = \mathcal{A}Q + 2\mathcal{C}Q^{[\mu\nu]}{}_{\nu}\lc{\nabla}_{\mu}\phi\,,
\end{equation}
and so the scalar field equation undergoes the trivial change to become
\begin{multline}
-(2\mathcal{A}\mathcal{B} + 3\mathcal{C}^2)\lc{\nabla}_{\mu}\lc{\nabla}^{\mu}\phi + (\mathcal{B}\mathcal{C} - 3\mathcal{C}\mathcal{C}' - \mathcal{A}\mathcal{B}')\lc{\nabla}_{\mu}\phi\lc{\nabla}^{\mu}\phi\\
+ (\mathcal{A}' + \mathcal{C})\left(\mathcal{A}Q + 2\mathcal{C}Q^{[\mu\nu]}{}_{\nu}\lc{\nabla}_{\mu}\phi\right) + 2\kappa^2(\mathcal{A}\mathcal{V}' + 2\mathcal{C}\mathcal{V}) = \kappa^2\mathcal{C}\Theta_{\mu}{}^{\mu}\,.
\end{multline}
This completes the field equations for the scalar-nonmetricity class of gravity theories.

We finally also take a brief look at the metric teleparallel case, and study the field equations of a class of scalar-torsion theories defined by the action~\cite{Geng:2011aj,Hohmann:2018rwf,Hohmann:2018ijr}
\begin{multline}\label{eq:sttaction}
S_{\text{g}} = \frac{1}{2\kappa^2}\int_M\Big[-\mathcal{A}(\phi)T - \mathcal{B}(\phi)g^{\mu\nu}\lc{\nabla}_{\mu}\phi\lc{\nabla}_{\nu}\phi\\
+ 2\mathcal{C}(\phi)T_{\nu}{}^{\nu\mu}\lc{\nabla}_{\mu}\phi - 2\kappa^2\mathcal{V}(\phi)\Big]\sqrt{-g}\dd^4x\,,
\end{multline}
which directly follows from the action~\eqref{eq:stgaction} by imposing vanishing nonmetricity. Recall that under this condition the single field equation obtained by simultaneous variation of the metric and connection is given by~\eqref{eq:metallfield}. Using the variation~\eqref{eq:stgvar}, these field equations become
\begin{multline}\label{eq:sttallfield}
\mathcal{A}\lc{R}_{\mu\nu} - \frac{\mathcal{A}}{2}\lc{R}g_{\mu\nu} + \mathcal{C}\lc{\nabla}_{\mu}\lc{\nabla}_{\nu}\phi - (\mathcal{B} - \mathcal{C}')\lc{\nabla}_{\mu}\phi\lc{\nabla}_{\nu}\phi + (\mathcal{A}' + \mathcal{C})S_{\mu\nu}{}^{\rho}\lc{\nabla}_{\rho}\phi\\
+ \left[\left(\frac{\mathcal{B}}{2} - \mathcal{C}'\right)\lc{\nabla}_{\rho}\phi\lc{\nabla}^{\rho}\phi - \mathcal{C}\lc{\nabla}_{\rho}\lc{\nabla}^{\rho}\phi + \kappa^2\mathcal{V}\right]g_{\mu\nu} = \kappa^2(\Theta^{\mu\nu} - \nabla_{\rho}H^{\mu\nu\rho} + H^{\mu\nu\rho}T^{\tau}{}_{\tau\rho})\,.
\end{multline}
These equations are supplemented by the scalar field equation, which follows from the general teleparallel equation~\eqref{eq:genscaldebfield} by using
\begin{equation}
\mathcal{A}G + 2\mathcal{C}M^{[\mu\nu]}{}_{\nu}\lc{\nabla}_{\mu}\phi = \mathcal{A}T - 2\mathcal{C}T_{\nu}{}^{\nu\mu}\lc{\nabla}_{\mu}\phi\,,
\end{equation}
in the absence of nonmetricity. Hence, the (debraided) scalar field equation takes the form
\begin{multline}
-(2\mathcal{A}\mathcal{B} + 3\mathcal{C}^2)\lc{\nabla}_{\mu}\lc{\nabla}^{\mu}\phi + (\mathcal{B}\mathcal{C} - 3\mathcal{C}\mathcal{C}' - \mathcal{A}\mathcal{B}')\lc{\nabla}_{\mu}\phi\lc{\nabla}^{\mu}\phi\\
+ (\mathcal{A}' + \mathcal{C})\left(\mathcal{A}T - 2\mathcal{C}T_{\nu}{}^{\nu\mu}\lc{\nabla}_{\mu}\phi\right) + 2\kappa^2(\mathcal{A}\mathcal{V}' + 2\mathcal{C}\mathcal{V}) = \kappa^2\mathcal{C}\Theta_{\mu}{}^{\mu}
\end{multline}
in the metric teleparallel gravity setting.

\subsection{Scalar-teleparallel representation of $f(G)$ theories}\label{ssec:stfg}
Among the general classes of scalar-teleparallel theories of gravity discussed in the previous section there is a particular subclass of theories, defined by a suitable choice of the parameter functions \(\mathcal{A}, \mathcal{B}, \mathcal{C}, \mathcal{V}\), whose field equations turn out to be equivalent to those of the \(f(G)\) class of theories. Note that for a given function \(f\), the choice of the parameter functions in the scalar-teleparallel representation is not unique, and different choices are connected by redefinitions of the scalar field. For the general teleparallel geometry, a straightforward procedure is to start from the action~\eqref{eq:fgaction}, and to rewrite it, similarly to the \(f(\lc{R})\) class of theories~\cite{Sotiriou:2008rp}, in the form
\begin{equation}\label{eq:stge2action}
S_{\text{g}} = -\frac{1}{2\kappa^2}\int_M[f(\phi) - \psi(\phi - G)]\sqrt{-g}\dd^4x\,,
\end{equation}
thereby introducing two scalar fields \(\psi\) and \(\phi\). Here  \(\psi\) is a Lagrange multiplier, and imposes the constraint
\begin{equation}\label{eq:stgcons1}
\phi = G
\end{equation}
for the scalar field \(\phi\). Variation with respect to the latter yields another constraint
\begin{equation}\label{eq:stgcons2}
\psi = f'(\phi)\,,
\end{equation}
which can then be used to solve for the scalar field \(\psi\). The remaining field equations are the metric field equation
\begin{multline}
\psi\left(\lc{R}_{\mu\nu} - \frac{1}{2}\lc{R}g_{\mu\nu}\right) + \left(\lc{\nabla}_{(\mu}\psi M^{\rho}{}_{\nu)\rho} - M^{\rho}{}_{(\mu\nu)}\lc{\nabla}_{\rho}\psi + M^{[\rho\sigma]}{}_{\sigma}\lc{\nabla}_{\rho}\psi g_{\mu\nu}\right)\\
+ \frac{1}{2}[f(\phi) - \phi\psi]g_{\mu\nu} = \kappa^2\Theta_{\mu\nu}\,,
\end{multline}
as well as the connection field equation
\begin{equation}
M^{\nu\rho}{}_{\rho}\lc{\nabla}_{\mu}\psi + M^{\rho}{}_{\mu\rho}\lc{\nabla}^{\nu}\psi - (M^{\nu\rho}{}_{\mu} + M^{\rho}{}_{\mu}{}^{\nu})\lc{\nabla}_{\rho}\psi = 2\kappa^2(\nabla_{\tau}H_{\mu}{}^{\nu\tau} - M^{\omega}{}_{\tau\omega}H_{\mu}{}^{\nu\tau})\,.
\end{equation}
Together with the constraints~\eqref{eq:stgcons1} and~\eqref{eq:stgcons2}, one finds that these reproduce the $f(G)$ field equations~\eqref{eq:fgmetfield2} and~\eqref{eq:fgconnfield2}.

Instead of keeping two scalar fields, one can take one further step and substitute the constraint~\eqref{eq:stgcons2} in the action~\eqref{eq:stge2action}, which then becomes
\begin{equation}\label{eq:stgeaction}
S_{\text{g}} = -\frac{1}{2\kappa^2}\int_M[f(\phi) - f'(\phi)(\phi - G)]\sqrt{-g}\dd^4x\,.
\end{equation}
Note that this does not change the metric and connection field equations. Variation with respect to the scalar field now yields the field equation
\begin{equation}
(G - \phi)f'' = 0\,,
\end{equation}
which resembles the constraint~\eqref{eq:stgcons1} for \(f'' \neq 0\). By comparison with the general scalar-teleparallel action~\eqref{eq:stgaction}, one reads off the relations
\begin{equation}\label{eq:stgeparfun}
\mathcal{A}(\phi) = f'(\phi)\,, \quad
\mathcal{B}(\phi) = 0\,, \quad
\mathcal{C}(\phi) = 0\,, \quad
\mathcal{V}(\phi) = \frac{f(\phi) - \phi f'(\phi)}{2\kappa^2}\,.
\end{equation}
Alternatively, if the constraint~\eqref{eq:stgcons2} is invertible, one may also solve it for \(\phi\) instead, which yields a different parametrization. The resulting action then takes the form
\begin{equation}\label{eq:stgeaction2}
S_{\text{g}} = -\frac{1}{2\kappa^2}\int_M[\psi G - 2\kappa^2\mathcal{U}(\psi)]\sqrt{-g}\dd^4x\,,
\end{equation}
where \(\mathcal{U}\) is implicitly defined by
\begin{equation}
\mathcal{U}(\psi) = \mathcal{V}(\phi)\,.
\end{equation}
To obtain a more explicit relation, one may differentiate with respect to \(\phi\) on both sides, which yields
\begin{equation}
f''(\phi)\mathcal{U}'(\psi) = -\frac{\phi f''(\phi)}{2\kappa^2}\,.
\end{equation}
This shows that \(f(\phi)\) and \(\mathcal{U}(\psi)\) are related by a Legendre transformation. In this case the scalar field equation becomes
\begin{equation}
G = -2\kappa^2\mathcal{U}'(\psi)\,,
\end{equation}
once again reproducing the constraint~\eqref{eq:stgcons1}, up to a change of parametrization.

It is easy to check that for the values~\eqref{eq:stgeparfun} of the parameter functions (and hence also for the equivalent parametrization via \(\psi\)) indeed yield a class of theories whose field equations reproduce those of the \(f(G)\), \(f(Q)\)~\cite{Jarv:2018bgs} and \(f(T)\)~\cite{Yang:2010ji} classes of gravity theories, if suitable restrictions are imposed on the torsion or nonmetricity of the connection. Substituting the values~\eqref{eq:stgeparfun} and the constraint~\eqref{eq:stgcons1} into the metric field equation~\eqref{eq:stgmetfield} yields
\begin{multline}
f'\lc{R}_{\mu\nu} - \frac{f'}{2}\lc{R}g_{\mu\nu} + f''\left(\lc{\nabla}_{(\mu}G M^{\rho}{}_{\nu)\rho} - M^{\rho}{}_{(\mu\nu)}\lc{\nabla}_{\rho}G + M^{[\rho\sigma]}{}_{\sigma}\lc{\nabla}_{\rho}G g_{\mu\nu}\right)\\
+ \frac{1}{2}(f - f'G)g_{\mu\nu} = \kappa^2\Theta_{\mu\nu}\,,
\end{multline}
which, using \(f''\lc{\nabla}_{\mu}G = \lc{\nabla}_{\mu}f'\), reproduces the field equation~\eqref{eq:fgmetfield2}. The same relation is used to show that the connection field equation~\eqref{eq:stgconnfield}, which becomes
\begin{multline}
f''\left[M^{\nu\rho}{}_{\rho}\lc{\nabla}_{\mu}G + M^{\rho}{}_{\mu\rho}\lc{\nabla}^{\nu}G - (M^{\nu\rho}{}_{\mu} + M^{\rho}{}_{\mu}{}^{\nu})\lc{\nabla}_{\rho}G\right]\\
= 2\kappa^2(\nabla_{\tau}H_{\mu}{}^{\nu\tau} - M^{\omega}{}_{\tau\omega}H_{\mu}{}^{\nu\tau})\,,
\end{multline}
resembles the connection field equation~\eqref{eq:fgconnfield2}. We then continue with the symmetric teleparallel case. Here, the connection equation~\eqref{eq:stqconnfield} becomes
\begin{equation}
-\nabla_{\nu}\nabla_{\rho}(f'\tilde{P}^{\nu\rho}{}_{\mu}) = \kappa^2\nabla_{\nu}\nabla_{\rho}\tilde{H}_{\mu}{}^{\nu\rho}\,,
\end{equation}
which is obviously identical to the corresponding equation~\eqref{eq:fqconnfield}. The metric field equation~\eqref{eq:stqmetfield} takes the form
\begin{equation}
f'\lc{R}_{\mu\nu} - \frac{f'}{2}\lc{R}g_{\mu\nu} - f''P^{\rho}{}_{\mu\nu}\lc{\nabla}_{\rho}Q + \frac{1}{2}(f - f'Q)g_{\mu\nu} = \kappa^2\Theta_{\mu\nu}\,.
\end{equation}
To bring this to the familiar form, one raises one index, and uses the fact that the left hand side of the STEGR field equation satisfies
\begin{equation}
-\frac{\nabla_{\rho}(\sqrt{-g}P^{\rho\mu}{}_{\nu})}{\sqrt{-g}} - \frac{1}{2}P^{\mu\rho\sigma}Q_{\nu\rho\sigma} + \frac{1}{2}Q\delta^{\mu}_{\nu} = R^{\mu}{}_{\nu} - \frac{1}{2}R\delta^{\mu}_{\nu}\,.
\end{equation}
This can be used to replace the Einstein tensor, so that the scalar-nonmetricity field equation becomes
\begin{equation}
-\frac{f'\nabla_{\rho}(\sqrt{-g}P^{\rho\mu}{}_{\nu})}{\sqrt{-g}} - \frac{f'}{2}P^{\mu\rho\sigma}Q_{\nu\rho\sigma} - P^{\rho\mu}{}_{\nu}\lc{\nabla}_{\rho}f' + \frac{1}{2}f\delta^{\mu}_{\nu} = \kappa^2\Theta^{\mu}{}_{\nu}\,.
\end{equation}
Observe that the two derivative terms can be combined into a single term, which yields the field equation~\eqref{eq:fqmetfield}. A similar procedure can be applied to the scalar-torsion case, whose field equations~\eqref{eq:sttallfield} now read
\begin{multline}
f'\lc{R}_{\mu\nu} - \frac{f'}{2}\lc{R}g_{\mu\nu} + f''S_{\mu\nu}{}^{\rho}\lc{\nabla}_{\rho}T + \frac{1}{2}(f - f'T)g_{\mu\nu}\\
= \kappa^2(\Theta^{\mu\nu} - \nabla_{\rho}H^{\mu\nu\rho} + H^{\mu\nu\rho}T^{\tau}{}_{\tau\rho})\,.
\end{multline}
Here one uses the left hand side of the MTEGR field equation, which can be written as
\begin{equation}
\lc{\nabla}_{\rho}(S_{\mu\nu}{}^{\rho}) + S^{\rho\sigma}{}_{\nu}K_{\rho\sigma\mu} + \frac{1}{2}Tg_{\mu\nu} = \lc{R}_{\mu\nu} - \frac{1}{2}\lc{R}g_{\mu\nu}\,.
\end{equation}
Using this relation to replace the Einstein tensor, one finds
\begin{multline}
f'\lc{\nabla}_{\rho}(S_{\mu\nu}{}^{\rho}) + f'S^{\rho\sigma}{}_{\nu}K_{\rho\sigma\mu} + S_{\mu\nu}{}^{\rho}\lc{\nabla}_{\rho}f' + \frac{1}{2}fg_{\mu\nu}\\
= \kappa^2(\Theta^{\mu\nu} - \nabla_{\rho}H^{\mu\nu\rho} + H^{\mu\nu\rho}T^{\tau}{}_{\tau\rho})\,.
\end{multline}
Once again the two derivative terms can be combined, and one has the field equation~\eqref{eq:ftallfield}.

\section{Outlook and open questions}\label{sec:outlook}
Teleparallel gravity theories are an active field of research and many questions are yet unanswered at the time of writing of this chapter. One of the most prominent open questions is known as the ``strong coupling problem''~\cite{BeltranJimenez:2019nns,Golovnev:2020zpv}. It refers to the fact that both the Hamiltonian analysis and higher order perturbation theory predict the presence of additional degrees of freedom compared to general relativity in several classes of teleparallel gravity, which are not manifest as propagating modes in the linear perturbation theory. Such modes are called strongly coupled, and their presence hints towards possible instabilities and a lack of predictability, which potentially renders the perturbation theory around such background solutions invalid. Among the most common approaches to clarify the nature and severity of these issues is the Hamiltonian analysis and the study of constraints.

Besides fundamental questions, also the phenomenology of teleparallel gravity theories leaves numerous possibilities for further studies, which can potentially lead to new experimental tests. Active fields at the time of writing this chapter include the study of cosmology using the method of dynamical systems, cosmological perturbations, black holes and other exotic compact objects, as well as their shadows and their perturbations, which are closely related to the emission of gravitational waves. Hence, it is reasonable to expect numerous future developments in this field.

\begin{acknowledgement}
The author thanks Claus Lämmerzahl and Christian Pfeifer for the kind invitation to contribute this book chapter. He acknowledges the full financial support of the Estonian Ministry for Education and Science through the Personal Research Funding Grant PRG356, as well as the European Regional Development Fund through the Center of Excellence TK133 ``The Dark Side of the Universe''.
\end{acknowledgement}

\bibliographystyle{spphys}
\bibliography{teleparallel}
\end{document}